\documentclass[a4paper,fleqn,usenatbib]{mnras}

\usepackage{graphicx}	% Including figure files
\usepackage{amsmath}	% Advanced maths commands
\usepackage{amssymb}	% Extra maths symbols
\usepackage{comment}

\newcommand{\rmem}{\overline{R}_{\rm{mem}}}
\newcommand{\redmap}{redMaPPer}
\newcommand{\redmapper}{redMaPPer}

\newcommand{\avg}[1]{\langle #1 \rangle}

\newcommand{\Mpc}{{\rm Mpc}}

\newcommand{\photoz}{photo$z$}

\newcommand\PhysRep{{Physics Reports}}

\defcitealias{Miyatake:2015}{M15}
\defcitealias{Simet:2016}{S16}

\title[\redmap{} Angular Clustering]{Constraining the Mass--Richness Relationship of \redmap{} Clusters with Angular Clustering}

\author[E. J. Baxter et al.]{
Eric J. Baxter,$^{1}$\thanks{E-mail: ebax@sas.upenn.edu}
Eduardo Rozo,$^{2}$
Bhuvnesh Jain,$^{1}$
Eli Rykoff,$^{3,4}$
Risa H. Wechsler$^{3,4}$
\\
$^{1}$Center for Particle Cosmology, Department of Physics, University of Pennsylvania, Philadelphia, PA 19104\\
$^{2}$Department of Physics, University of Arizona, Tucson, AZ 85721, USA \\
$^{3}$Kavli Institute for Particle Astrophysics and Cosmology, Department of Physics, Stanford University, Stanford, CA, 94305 \\
$^{4}$SLAC National Accelerator Laboratory, Menlo Park, CA, 94025
}

\date{Last updated \today}

\pubyear{2016}

\begin{document}
\label{firstpage}
\pagerange{\pageref{firstpage}--\pageref{lastpage}}
\maketitle

% Abstract of the paper
\begin{abstract}
The potential of using cluster clustering for calibrating the
mass-observable relation of galaxy clusters has been recognized
theoretically for over a decade.  Here, we demonstrate the feasibility
of this technique to achieve high precision mass calibration using
\redmap{} clusters in the Sloan Digital Sky Survey North Galactic Cap.
By including cross-correlations between several richness bins in our
analysis we significantly improve the statistical precision of our
mass constraints.  The amplitude of the mass--richness relation is
constrained to $7\%$ statistical precision.  However, the error budget
is systematics dominated, reaching an 18\% total error that is
dominated by theoretical uncertainty in the bias--mass relation for
dark matter halos.  We perform a detailed treatment of the effects of
assembly bias on our analysis, finding that the contribution of such
effects to our parameter uncertainties is somewhat greater than that
of measurement noise.  We confirm the results from
\citet{Miyatake:2015} that the clustering amplitude of \redmapper{}
clusters depends on galaxy concentration, and provide additional
evidence in support of this effect being due to some form of assembly
bias.  The results presented here demonstrate the power of cluster
clustering for mass calibration and cosmology provided the current
theoretical systematics can be ameliorated.
\end{abstract}


%%%%%%%%%%%%%%%%%%%%%%%%%%%%%%%%%%%%%%%%%%%%%%%%%%

%%%%%%%%%%%%%%%%% BODY OF PAPER %%%%%%%%%%%%%%%%%%

\section{Introduction}
\label{sec:intro}

The abundance of galaxy clusters is a powerful probe of cosmology
\citep{weinbergetal13}.  The observed number density of clusters as a
function of redshift, $n(z)$, is sensitive to both the expansion
history of the Universe and to the growth of structure.  This dual
sensitivity enables cluster abundance measurements to distinguish
between dark energy models and modified gravity \citep{Huterer:2013}.
However, modeling the cluster abundance to extract cosmological
constraints is challenging because the abundance depends sensitively
on the masses of the observed clusters.  Because clusters live at the
extreme tails of the mass function where the abundance is falling
exponentially, small changes in the cluster mass can have a large
impact on the predicted cluster abundance.  Moreover, measuring
cluster masses is difficult because it requires relating observable
quantities to cluster mass.  The mass--observable relationships are
often noisy and subject to large systematic uncertainties
\citep{rozoetal14a,serenoetal15}.  For these reasons, the dominant
systematic currently affecting cosmological constraints derived from
cluster abundance measurements is uncertainty in cluster masses
\citep{Vikhlinin:2009, Rozo:2010, Sehgal:2011,
  Benson:2013,mantzetal15,planck_clusters15}.  If these mass
uncertainties can be reduced, the full power of cluster abundance
measurements can be exploited, and clusters will become highly
competitive probes of dark energy and modified gravity
\citep{Albrecht:2006}.

The spatial clustering of galaxy clusters is sensitive to cluster
mass, with higher mass clusters being more strongly clustered on the
sky.  To quantify the degree of clustering it is common to define the
{\it bias} as the square root of the ratio of the correlation function
of clusters --- which we identify with massive dark matter halos ---
to that of all matter.  The relationship between halo bias and halo
mass, $b(M)$, is well understood via the halo model \citep[for a
  review see][]{Cooray:2002} and can be calibrated from simulations
\citep[e.g.][]{Tinker:2010}.  Several authors have highlighted the
possibility of using measurements of the clustering biases of clusters
to do a ``self-calibrated'' cluster cosmology analysis, i.e. to
jointly fit for cosmological parameters as well as parameters governing
the relationship between observable quantities and cluster mass
\citep[e.g.][]{Majumdar:2003, Lima:2004, Hu:2006, Holder:2006}.

In this work, we perform the first half of the self-calibration
program, i.e. measuring the clustering biases of a catalog of
optically detected galaxy clusters and using the measured biases to
constrain the mass--observable relationship.  To this end, we use the
catalog of \redmap{} clusters \citep{Rykoff:2014} identified in the
Sloan Digital Sky Survey (SDSS).  The relevant observable for the
\redmap{} clusters is the {\it richness}, a measure of the number of
galaxies within the cluster. Using the measured correlation functions
and the predicted $b(M)$, we calibrate the mass--richness relation for
the \redmap{} clusters.

Two of the most significant challenges for using cluster clustering to
calibrate cluster masses are line-of-sight cluster projections and
assembly bias.  Chance alignment of clusters along the line of sight
can alter the richness estimates of these clusters and change the
measured clustering amplitude.  If this effect is not accounted for,
constraints on the mass--richness relation of the \redmap{} clusters
could be biased.  Assembly bias refers to the dependence of halo
clustering on assembly history or other quantities in addition to halo
mass \citep{Gao:2005,Wechsler:2006,Jing:2007,Dalal:2008}.  Recently,
\citet{Miyatake:2015} have measured dependence of the clustering of
\redmap{} clusters on $\rmem$, the average radial separation between
the cluster center and its member galaxies.  Dependence of the
clustering amplitude on $\rmem$ (or any other properties of the halo
beyond the halo mass dependence) complicates any attempt to extract
constraints on the mass--richness relation or cosmology using $b(M)$.
Below, we develop methodology to account for the effects of both
cluster projections and assembly bias on our constraints on the
parameters of the mass--richness relationship.

The outline of the paper is as follows: we introduce the formalism
for describing the correlation function measurements in
\S\ref{sec:formalism}; the cluster catalog is described in
\S\ref{sec:data}; the measurement of the correlation function and its
covariance are described in \S\ref{sec:measurement}; our model for the
correlation function and fitting procedure --- including our treatment
of projection and assembly bias effects --- are described in
\S\ref{sec:analysis}; results are presented in \S\ref{sec:results}; we
explore the various contributions to our error budget in
\S\ref{sec:systematics}; the cosmology dependence of our constraints
is considered in \S\ref{sec:cosmology_dependence}; finally, our
conclusions are given in \S\ref{sec:discussion}.

Throughout, all cluster masses refer to $M_{200m}$, i.e. the mass
enclosed within a sphere centered on a cluster such that the mean
density within that sphere is 200 times the mean density of the
Universe at the cluster's redshift.  Our fiducial analysis assumes the
best-fit flat $\Lambda$CDM model from an analysis of cosmic microwave
background data and other data sets by \citet{PlanckXIII:2015}. The
cosmological parameters in this model are $\Omega_m = 0.309$, $h_0 =
H_0/(100{\rm km/s/Mpc}) = 0.677$, $\Omega_b = 0.0486$, $\tau = 0.066$,
$n_s = 0.9667$ and $A_s = 2.14\times 10^{-9}$ at a pivot scale of $k =
0.05\, {\rm Mpc}^{-1}$.  In \S\ref{sec:cosmology_dependence} we
explore how our parameter constraints are affected by variations in
the assumed cosmological model.

%%%%%%%%%%%%%%%%%%%%%%%%%%%%%%%
%%%%%%%%%%%%%%%%%%%%%%%%%%%%%%%
%%%%%%%%%%%%%%%%%%%%%%%%%%%%%%%
%%%%%%%%%%%%%%%%%%%%%%%%%%%%%%%
%%%%%%%%%%%%%%%%%%%%%%%%%%%%%%%
%%%%%%%%%%%%%%%%%%%%%%%%%%%%%%%
%%%%%%%%%%%%%%%%%%%%%%%%%%%%%%%
%%%%%%%%%%%%%%%%%%%%%%%%%%%%%%%

\section{Formalism}
\label{sec:formalism}

We define $n(\hat{\phi})$ to be the projected density of clusters in
the direction specified by the unit vector $\hat{\phi}$.  The
overdensity in the same direction is then defined as
\begin{eqnarray}
\delta(\hat{\phi}) \equiv \frac{n(\hat{\phi}) -
\bar{n}(\hat{\phi})}{\bar{n}(\hat{\phi})},
\end{eqnarray}
where $\bar{n}(\hat{\phi})$ is the average of $n(\hat{\phi})$ over all
$\hat{\phi}$.  The clustering of clusters on the sky can be characterized in
terms of the angular correlation function, $w(\theta)$:
\begin{eqnarray}
w(\theta) = \langle \delta(\hat{\phi}) \delta(\hat{\phi}')\rangle,
\end{eqnarray}
where the average is taken over all possible $\hat{\phi}$ and
$\hat{\phi'}$ such that the angular separation between $\hat{\phi}$
and $\hat{\phi}'$ is $\theta$.

We measure the angular correlation function of the \redmap{} clusters
in bins of richness and redshift; in addition, we also measure several
cross-bin angular correlations. To keep the notation simple, we use a
single Greek superscript to represent both the richness and redshift
bins.  The correlation function between richness/redshift bin $\alpha$
and richness/redshift bin $\beta$ will be denoted with
$w^{\alpha\beta}(\theta)$.  Occasionally we suppress the superscripts
on $w(\theta)$ for notational convenience.

The correlation function $w^{\alpha\beta}(\theta)$ is related to the
angular power spectrum, $C_{\ell}^{\alpha\beta}$, by
\begin{eqnarray}
\label{eq:wtheta}
w^{\alpha\beta}(\theta) = \sum_{l= 0}^{\infty} \left( \frac{2l + 1}{4\pi}\right) P_{\ell}(\cos \theta) C_{\ell}^{\alpha\beta},
\end{eqnarray}
where $P_{\ell}$ is the Legendre polynomial of order $\ell$.  We can relate
$C_{\ell}^{\alpha\beta}$ for \redmap{} clusters to the matter power spectrum using
the Limber approximation:
\begin{eqnarray}
\label{eq:limber}
C_{\ell}^{\alpha\beta} = \int dz\ W^{\alpha}(z) W^{\beta}(z) \frac{H(z)}{d^2_A(z)}b^2(k,z) P\left( k = \frac{\ell+1/2}{\chi(z)};z \right),
\end{eqnarray}
where $H(z)$ is the Hubble parameter at redshift $z$, $d_A(z)$ is the
angular diameter distance, $\chi(z)$ is the comoving distance, and we
have assumed a spatially flat Universe \citep{Limber:1953}.  The
weight function $W^{\alpha}(z)$ is the distribution of clusters in the
$\alpha$-th richness/redshift bin as a function of redshift,
normalized such that $\int dz\ W^{\alpha}(z) = 1$.  The $b(z,k)$ term
is the bias of the \redmap{} clusters, which we discuss in more
detail below.  The Limber approximation is expected to be valid when
the redshift selection function, $W^{\alpha}(z)$, is much wider than
the scales of interest.  In our case, the maximum scale that we probe
is $\sim 25\,{\rm Mpc}$, while the width of the selection function is
$\sim 280\,{\rm Mpc}$.  We are therefore safely in the regime for
which the Limber approximation should hold.

At large scales, the bias is expected to be independent of $k$.  This
is the so-called linear bias regime.  At small scales, the bias may
become scale-dependent and is difficult to model.  We therefore
restrict our analysis to the linear bias regime (but still at scales
small enough that the Limber approximation holds). In this limit, the
angular correlation function is simply
\begin{eqnarray}
w^{\alpha\beta} (\theta) = b^{\alpha} b^{\beta}
w_M^{\alpha\beta}(\theta),
\end{eqnarray}
where $w_M^{\alpha\beta}$ is the matter-matter correlation function,
given by Eqs.~\ref{eq:wtheta} and \ref{eq:limber} with $b(k,z) = 1$.
The coefficients $b^{\alpha}$ and $b^{\beta}$ are the linear bias
parameters for the two richness/redshift bins $\alpha$ and $\beta$;
constraining these parameters for the \redmap{} clusters is one of the
main goals of this analysis. The bias parameters should be thought of
as averages across all clusters in a given richness/redshift bin.

\section{Data}
\label{sec:data}

Our analysis uses the \redmap{} \citep{Rykoff:2014} catalog of
clusters identified in the SDSS \citep[SDSS:][]{York:2000} $8^{th}$
Data Release \citep[DR8][]{Aihara:2011}.  The SDSS DR8 photometric
galaxy catalog includes roughly $14,000\ \deg^2$ of imaging, which was
reduced to $\sim 10,000\ \deg^2$ of high quality contiguous imaging
for the purposes of cluster finding using the Baryon Oscillation
Spectroscopic Survey (BOSS) mask \citep{Dawson:2013}.  Given a cluster
candidate, \redmap{} uses the SDSS 5-band imaging to estimate the
probability of any given galaxy in the field of being a red-sequence
cluster member of the candidate cluster.  The cluster richness is the
total number of red-sequence members $\lambda$, and serves as our
observable mass proxy.  A detailed explanation of the \redmap{}
algorithm can be found in \citet{Rykoff:2014}.

For the main results of this work we restrict our analysis to
\redmap{} clusters identified in the North Galactic Cap (NGC) of SDSS,
a contiguous region of $\sim 7,000\ \deg^2$ in the northern
hemisphere.  We restrict our analysis to the NGC for two reasons.
First, the weak lensing mass calibration of \redmap{} clusters
performed by \citet{Simet:2016} (hereafter \citetalias{Simet:2016}) is
restricted to the NGC (for reasons discussed therein) and imposing the
same restriction on our analysis makes comparison between the two
works more straightforward. Second, as we discuss in
\S\ref{subsec:err_sysweights}, we find some tension between the
clustering measurements in the NGC and the Southern Galactic Cap
(SGC).  We refrain from combining the parameter constraints derived
from the NGC and SGC because of this tension, and only present results
from the NGC because it represents a larger area than the SGC.

Our analysis relies on the \redmap{} v5.10
catalog,\footnote{\texttt{http://risa.stanford.edu/redmapper/}} an
updated version of the original \redmap{} cluster catalog with a
variety of modest improvements.  Typical photometric redshift
uncertainties are $\sigma_z/(1+z) \lesssim 0.01$, with \photoz\ biases
controlled at the $\Delta z \approx 0.003$ level. This level of
photometric redshift performance is sufficiently high that photometric
redshift errors can be safely ignored in our analysis.

We note that while the \redmap\ v5.10 cluster catalog is publicly
available, our analysis relies on a proprietary version of the catalog
that extends the low richness threshold from $\lambda\geq 20$ to
$\lambda\geq 5$. As discussed in \citet{Rykoff:2014}, the $\lambda\geq
20$ threshold is a purposely conservative selection that ensures a
clean connection between individual dark matter halos and galaxy
clusters selected by \redmap.  We demonstrate below why our specific
science goal allows us to use low richness clusters, even if the
connection between halos and galaxy clusters is less secure for these
low richness systems.  

We restrict the catalog used in this work to clusters with $0.18 < z <
0.33$.  The cut at the high redshift end ensures that the cluster
catalog is volume limited \citep{Rykoff:2014}.  The cut at the low
redshift end is chosen to maximize our signal to noise while ensuring
that we remain in the linear bias regime (see discussion in
\S\ref{sec:measurement}).

\section{Correlation Function Measurement}
\label{sec:measurement}

We measure $w^{\alpha\beta}(\theta)$ using the pair counting estimator
of \citet{Landy:1993}.  The \citet{Landy:1993} estimator relies on
computing three quantities to estimate the correlation function over a
given angular bin: $DD$, $RR$, and $DR$.  The quantity $DD$
(`data-data') is the number of pairs of clusters that have an angular
separation that is within the bin, $RR$ (`random-random') is similarly
defined for a catalog of points whose positions have been randomly and
independently chosen on the sky, and $DR$ (`data-random') is the
number of cross pairs between the cluster and random catalogs.  To
reduce the effects of statistical noise, the random catalog is
typically generated with $\sim 30$ to $50$ times more points than the
data catalog (although, more optimal estimators such as that of
\citet{Baxter:2013} can reduce this requirement).

For a particular angular bin, the \citet{Landy:1993} estimate of the
correlation function between richness/redshift bin $\alpha$ and
richness/redshift bin $\beta$ is
\begin{multline}
\label{eq:ls_estimator}
\hat{w}^{\alpha\beta} =
\left(\frac{DD^{\alpha\beta}}{RR^{\alpha\beta}}\right)\left(\frac{R^\alpha
  R^\beta}{D^\alpha D^\beta} \right) -
\left(\frac{DR^{\alpha\beta}}{RR^{\alpha\beta}}\right)\left(\frac{R^{\alpha}}{D^{\alpha}} \right) \\ -
\left(\frac{DR^{\beta\alpha}}{RR^{\alpha\beta}}\right)\left(\frac{R^{\beta}}{D^\beta}
\right) + 1,
\end{multline}
where $D^{\alpha}$ and $R^{\alpha}$ are the number of data and random
points in the $\alpha$ richness/redshift bin, respectively (and
similarly for $D^{\beta}$ and $R^{\beta}$).  The superscript notation
indicates which bins are being correlated; $DD^{\alpha\beta}$, for
instance, refers to the pair counts between bins $\alpha$ and $\beta$.
All pair counting in our analysis is performed using the tree code
\verb!TreeCorr! \footnote{\texttt{https://github.com/rmjarvis/TreeCorr}}
\citep{Jarvis:2004}.

In order to account for survey geometry and selection effects, the
random catalog used to compute $DR$ and $RR$ must be restricted to the
same sky mask as the data.  As the \redmap{} mask is richness and
redshift dependent, the creation of an appropriate random catalog is
non-trivial.  The random catalog used in this work is generated as
follows: for each cluster in the \redmap{} catalog, we select a random
position in the sky, and test whether the cluster could have been
detected at that location.  If yes, the cluster is added to the random
point catalog at this random location.  If not, that cluster is
rejected from the random catalog.  Because of the large number of
draws, every cluster is drawn multiple times, and we record the number
of times the cluster was added to the random catalog ($N_1$), and the
number of times the cluster was rejected ($N_2$).  Every cluster is
assigned a weight $W=(N_1+N_2)/N_1$ in the random catalog to ensure
that the weighted random points statistically reproduce the joint
richness, redshift, and spatial distribution of the parent sample.  We
modify Eq.~\ref{eq:ls_estimator} so as to include these weights.
Specifically, a random point with weight $W_i$ contributes $W_i$
to $R$ and $DR$, and two random points $i$ and $j$ contributes a
weight $W_iW_j$ to $RR$.

Our baseline analysis uses $N_{\theta} = 4$ angular bins, $N_{\lambda}
= 4$ richness bins and $N_{z} = 2$ redshift bins.  We discuss the
choice of angular bins in more detail in
\S\ref{subsec:angular_binning}.  The richness bin edges are
$[5,20,28,41, \infty)$, while the redshift bin edges are
$[0.18,0.26,0.32]$.  The bins have been chosen so that the measurement
of $w^{\alpha\beta}(\theta)$ in each richness and redshift bin has
roughly the same signal-to-noise.

In addition to measuring the angular auto-correlation functions of the
clusters in each richness and redshift bin, we also measure the
angular cross-correlations of clusters in the same redshift bin, but
in different richness bins.  Including these cross-correlations
considerably enhances the signal-to-noise of our constraints on the
clustering biases.  Because $\sigma_z$ is small for \redmap{} clusters
($\sigma_z \lesssim 0.01$) and because the redshift bins are wide
($\Delta z \sim 0.07$), the angular cross-correlations between
different redshift bins are expected to be negligible and therefore
contribute little information to our bias constraints.  Since
including these measurements would come at the cost of increasing the
size of the covariance matrix --- and would therefore make our
jackknife covariance estimation less reliable (see below) --- we do
not include measurements of cross-redshift bin correlations as part of
our analysis.  We have, however, confirmed that these correlations are
negligible.  For our baseline analysis the data vector has dimension
$d = N_{\theta}N_{\lambda}(N_{\lambda} + 1)N_{z}/2 = 80$.

We compute the covariance matrix of our $w(\theta)$ measurements using
a jackknife approach.  To estimate the covariance matrix with a
jackknife, the survey region is divided into $N_{{\rm jk}}$ patches.
The measurement of $w(\theta)$ is repeated with each of the patches
removed in turn.  We use $\vec{w}_i$ to denote the estimate of
$w(\theta)$ with the $i$-th patch removed, where the vector notation
indicates that $\vec{w}_i$ includes measurements at several $\theta$
values and for several richness/redshift bins ($\vec{w}_i$ is assumed
to be a column vector).  The jackknife covariance matrix estimate is
\begin{eqnarray}
\hat{\mathbf{C}} = \frac{N_{\rm jk} - 1}{N_{\rm jk}} \sum_i^{N_{\rm jk}} \left( \vec{w_i} - \vec{\bar{w}} \right) \left( \vec{w}_i - \vec{\bar{w}} \right)^T,
\end{eqnarray}
where 
\begin{eqnarray}
\bar{w}(\theta) = \frac{1}{N_{\rm jk}} \sum_{i=1}^{N_{\rm jk}} \vec{w}_i(\theta)
\end{eqnarray}
\citep[see e.g.][]{Norberg:2009}.  In this work we use $N_{\rm jk}
\sim 800$ patches whose edges are along lines of constant right
ascension and declination.  Since the data vector has 80 elements,
there are roughly 10 times as many jackknife patches as the dimension
of the covariance matrix.

\subsection{Angular Binning}
\label{subsec:angular_binning}

In principle, since all angular scales contain information about
$w(\theta)$, one would like to measure $w(\theta)$ across the widest
possible range of $\theta$ to extract the tightest possible
constraints on the biases of \redmap{} clusters.  At low $\theta$,
however, modeling $w(\theta)$ becomes increasingly difficult for two
reasons.  First, modeling the matter power spectrum at small scales is
difficult because of baryonic effects \citep{Jing:2006,
  vanDaalen:2011}.  Second, at small scales the bias is expected to
become scale-dependent, and modeling this scale dependence is
difficult \citep[see e.g.][for discussion within the context
of halo-mass correlation functions]{hayashiwhite08,zuetal12}. 

To eliminate modeling uncertainties at small angular scales, we
restrict our analysis to the angular regime for which the power
spectrum can be accurately modeled and the linear bias model is
expected to be valid.  Imposing this requirement greatly simplifies
the analysis as it leaves only one free parameter --- the linear bias,
$b^{\alpha}$ --- to describe the clustering of \redmap{} clusters in
the $\alpha$ richness/redshift bin (given a fixed cosmological model).

\citet{zuetal12} find that a linear bias model can adequately describe
the halo-mass correlation function for scales $R\gtrsim 3.5\ \Mpc$.
Being conservative, we therefore set $\theta_{{\rm min}}$ by demanding
that the projected distance $R$ spanned by the angle $\theta_{\rm
  min}$ be more than a minimum length $R_{\rm min}=5\ h^{-1}\Mpc$ for
all clusters in a given redshift bin.  This amounts to setting
$\theta_{{\rm min}} \approx 39'$ for the first redshift bin and
$\theta_{{\rm min}} \approx 30'$ for the second redshift bin.  The
boundary of the lowest redshift bin has been chosen to maximize
signal-to-noise while maintaining the requirement that $\theta_{{\rm
    min}} d_A(z_{{\rm min}}) = 5\ h^{-1}{\rm Mpc}$.  We emphasize that
the minimum cutoff length is especially conservative in that the
cutoff distance is a distance {\it in projection}; the true
three-dimensional separation is expected to be significantly larger
($\approx 25\ \Mpc$ for our cluster sample).

At large scales, modeling the $w^{\alpha\beta}(\theta)$ is
straightforward, but computing the covariance of
$\hat{w}^{\alpha\beta}(\theta)$ with the jackknife becomes difficult.
The jackknife covariance estimate is only expected to yield reliable
results for angular scales less than the size of the jackknife
patches.  For our $N_{\rm jk} \sim 800$ patches, the typical size of a
patch is $\theta_{\rm patch} \sim 170'$.  We therefore set
$\theta_{\rm max} = 100'$, roughly a factor of 1.7 smaller than the
size of the jackknife patches.  Our four angular bins are spaced
logarithmically between $\theta_{\rm min}$ and $\theta_{\rm max}$.

\begin{figure}
\includegraphics[width = \columnwidth]{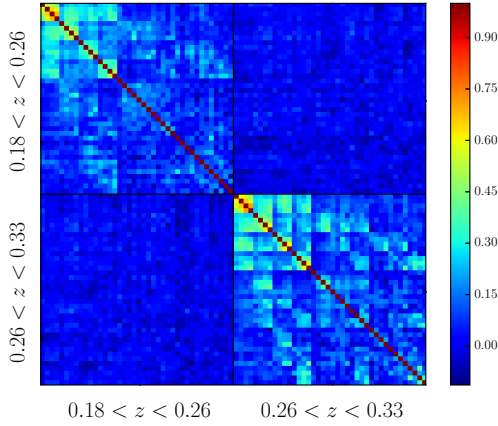}
\caption{The full correlation matrix
  $\mbox{Cov}(X_i,X_j)/\sigma_i\sigma_j$ between each measured $X_i =
  \hat{w}^{\alpha\beta}(\theta)$, computed using a spatial jackknife
  with roughly 800 subregions.  See text for the ordering of the
  matrix elements.  Note that the clustering signals from clusters in
  different redshift bins are uncorrelated.}
\label{fig:corr_mat}
\end{figure}

Fig.~\ref{fig:corr_mat} shows the correlation matrix for the data
vector as measured using the jackknife with roughly 800 subregions.  The
data vector is ordered such that the first 40 elements correspond to
the first redshift bin and the last 40 correspond to the second
redshift bin.  For each redshift bin, the elements are ordered
following the arrangement in Fig.~\ref{fig:data_z1} and
Fig.~\ref{fig:data_z2}: the first 16 elements correspond to the
measurements in the first row of those figures, the next 12 elements
correspond to the second row, etc.  The key take away from this figure
is that there is little covariance between clusters in different
redshift bins, but that there is significant covariance between
different richness bins within the same redshift slice.

\subsection{Correction for Observational Systematics}
\label{subsec:sys_correction}

If the density of galaxy clusters is modulated by observing
conditions, the variation in observing conditions across the survey
will induce artificial power in the clustering of clusters.  To
correct for this potential source of systematic error, we evaluate
several measures of observation quality at each cluster's location ---
including sky flux, dust extinction, size of the point spread function
and survey depth --- and then bin the clusters by these measures.  We
then determine the area sampled by each bin, and calculate the
corresponding cluster number density.  In the absence of observational
systematics, we expect the cluster density to be independent of the
various measures of observation quality.

\begin{figure}
  \includegraphics[width =\columnwidth]{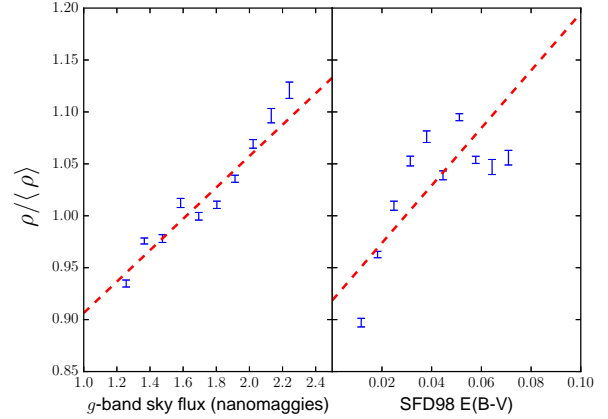}
  \caption{The measured variation in \redmap{} cluster density with
    $g$-band sky flux and $E(B-V)$ dust extinction measured by
    \citet{Schlegel:1998}.  Dashed lines show the linear fits that we
    use to correct our clustering measurements for these variations.}
  \label{fig:density_vs_systematic}
\end{figure}

Fig.~\ref{fig:density_vs_systematic} shows how the \redmapper{}
cluster density varies with both sky flux in the $g$-band (left panel)
and $E(B-V)$ dust extinction measured by \citet{Schlegel:1998} (right
panel).  We have selected these two potential sources of systematic
contamination because they exhibit the strongest effect on the
\redmapper\ cluster density.  Variations in $g$-band sky flux and dust
extinction can lead to $\sim 10\%$ changes in the cluster density.  In
order to correct our analysis for these effects, we introduce a
systematic weight, $w_i^{\rm sys}$, for the $i$-th cluster:
\begin{eqnarray}
w_i^{\rm sys} = \frac{ A + Bs_i}{\bar n},
\end{eqnarray}
where $\bar n$ is the average cluster density, $s_i$ is the measure of
the observational systematic evaluated at the $i$-th cluster, and $A +
Bs_i$ is the linear fit to the observed variations in cluster density
shown as the dashed lines in Fig.~\ref{fig:density_vs_systematic}.
The weight $w_i^{\rm sys}$ is the relative over-sampling of regions
that have observation quality described by $s_i$.  Consequently, we
can correct for this systematic by applying $w_i^{\rm sys}$ to each
random point.  We choose to only weight the random catalog by sky flux
in the $g$-band as the correlation between this systematic and the
\redmap{} cluster density appears particularly tight.  In
\S\ref{subsec:err_sysweights} we determine the level of systematic
uncertainty introduced into our parameter constraints by this choice.

\section{Analysis}
\label{sec:analysis}

Given bias parameters $b^{\alpha}$ for each richness/redshift bin and
a cosmological model, we can compute the predicted
$w^{\alpha\beta}(\theta)$ using Eqs.~\ref{eq:wtheta} and
\ref{eq:limber}.  We compute the matter power spectrum, $P(k,z)$, in
Eq.~\ref{eq:limber} using CAMB\footnote{\texttt{http://camb.info}}
\citep{Lewis:1999} with the Halofit prescription for the non-linear
power spectrum \citep{Smith:2003} including the updates of
\citet{Takahashi:2012}.  At scales larger than $\sim\ 5\ {\rm Mpc}$
the influence of baryons on the matter power spectrum is expected to
be at most a few percent \citep[e.g.][]{vanDaalen:2011}.  Since our
analysis only uses scales larger than $5\ h^{-1} {\rm Mpc}$, the
effects of baryons on the matter power spectrum are well below the
statistical uncertainty of our measurements.  The accuracy of Halofit
should therefore be sufficient for this analysis.

\subsection{Bias-only Model}
\label{subsec:biasonly_param}

One way to model the bias, $b^{\alpha}$, of each bin of clusters is to
simply treat the $b^{\alpha}$ as free parameters.  The results of
fitting such a model to the observed cluster correlation functions are
presented in \S\ref{sec:results}.  While constraints on the
$b^{\alpha}$ are potentially useful for a cosmological analysis of
\redmap{} clusters, they suffer from the drawback that the
$b^{\alpha}$ depend on the richness and redshift bin choices.

An alternate and bin-independent approach is to parameterize the
relationship between cluster bias and the observed richness and
redshift of each cluster.  This parameterization can then be
constrained by fitting to the measured angular correlation function.
We adopt a simple model for the richness and redshift dependence of
the bias:
\begin{eqnarray}
\label{eq:bias_vs_lam_z}
b(\lambda, z) = A_{\lambda}\left(\frac{\lambda}{\lambda_0}\right)^{\alpha_{\lambda}}\left( \frac{1+z}{1+z_0}\right)^{\beta_{\lambda}}
\end{eqnarray}
where $A_{\lambda}$, $\alpha_{\lambda}$ and $\beta_{\lambda}$ are free
parameters.  We set $\lambda_0 = 35$ and $z_0 = 0.25$ throughout;
these values are chosen because they are roughly the median richness
and redshift for our sample, respectively.  

When constraining the bias-richness parameterization and the
mass--richness relation of \redmap\ clusters, we restrict the
application of these parameterizations to clusters with $\lambda \geq
20$.  As discussed in \citet{Rykoff:2014}, our {\it a priori}
expectation is that the connection between galaxy clusters and
individual dark matter halos is very clean at high richness, but may
become more problematic with decreasing richness.  The richness
threshold $\lambda \geq 20$ is one we believe to be very conservative.
However, the lowest richness clusters with richness $5 < \lambda < 20$
are still expected to be perfectly good mass tracers.  Consequently,
we model the bias of the richness bin $5 < \lambda<20$ with a single
bias parameter, $b^0$ (with each redshift bin having a different $b^0$
parameter).  Because the cross-correlations between the $5 < \lambda <
20$ clusters and the $\lambda \geq 20$ clusters contain information
about the biases of the $\lambda \geq 20$ clusters, and because there
are many clusters with $5 < \lambda < 20$, including these
cross-correlations in our analysis significantly reduces the
statistical error of our bias measurements.

For the $\lambda \geq 20$ clusters, Eq.~\ref{eq:bias_vs_lam_z} yields
the bias for a cluster at a particular richness and redshift.  What we
measure, however, is the bias averaged over a richness/redshift bin.
To model the bias of a richness/redshift bin we simply average
Eq.~\ref{eq:bias_vs_lam_z} over the bin:
\begin{eqnarray}
\label{eq:bias_avg}
b^{\alpha} = \frac{1}{N_{\alpha}}\sum_i^{N_{\alpha}} b(\lambda_i,z_i),
\end{eqnarray}
where the sum runs over the $N_{\alpha}$ clusters in the
richness/redshift bin $\alpha$.  Our model therefore takes into
account the richness and redshift distribution of all of the clusters
in the bin.  Note that at this stage, our modeling is completely
independent of the mass--richness relation.  

\subsection{Mass--observable Parameterization}

Above, we parameterized the bias of the \redmap{} clusters in terms of
the cluster richness and redshift.  Alternatively, if both the
mass--bias relationship and the mass--richness relation for
\redmap\ clusters is known, one can use these relations to predict the
clustering bias --- and therefore the angular correlation function ---
of the cluster population.  Parameterizing in this way allows one to
use the data to constrain the mass--richness model for the \redmap{}
clusters.  

For clusters with $\lambda \geq 20$, we model the bias as a function
of richness and redshift with
\begin{eqnarray}
\label{eq:bias_vs_m_z}
b(\lambda, z) = \int d\ln M \, b(\ln M, z) P(\ln M|\lambda, z), 
\end{eqnarray}
where $P(\ln M|\lambda, z)$ describes the probability of a cluster
having log-mass $\ln M$ given that it has observed richness $\lambda$
and redshift $z$.  The term $b(\ln M,z)$ is the bias of a cluster with
log-mass $\ln M$ and redshift $z$.  We adopt the best-fit model for
$b(\ln M,z)$ from \citet{Tinker:2010}, where $b(\ln M,z)$ was
calibrated using N-body simulations.

We consider a Gaussian model for $P(\ln M | \lambda, z)$:
\begin{eqnarray}
\label{eq:P_of_lnM}
P(\ln M | \lambda, z) &=& \frac{1}{\sqrt{2\pi \sigma_{\ln M}^2 }} \exp \left[ -\frac{\left( \ln M - \left< \ln M | \lambda, z \right>\right)^2 }{2\sigma_{\ln M}^2}\right]. \nonumber \\
\end{eqnarray}
To parameterize $\left< M | \lambda, z \right>$ we adopt the model
\begin{eqnarray}
\label{eq:meanM}
\left< M | \lambda, z \right> = A ( \lambda/\lambda_0)^{\alpha} \left(  \frac{1+z}{1+z_0} \right)^{\beta}.
\end{eqnarray}
We treat $\sigma_{\ln M}$, $\ln A$, $\alpha$, and $\beta$ as free
parameters.  This parameterization is equivalent to
\begin{eqnarray}
\label{eq:meanlnM}
 \left< \ln M | \lambda \right> = \ln A + \alpha\ln( \lambda/\lambda_0) + \beta\ln\left(  \frac{1+z}{1+z_0} \right) -  \frac{1}{2}\sigma_{\ln M}^2,
\end{eqnarray}
which can be directly substituted into
Eq.~\ref{eq:P_of_lnM}.\footnote{We have also considered the alternate
  (but similar) parameterization $\left< \ln M | \lambda \right> = \ln
  A + \alpha\ln( \lambda/\lambda_0) + \beta\ln\left( \frac{1+z}{1+z_0}
  \right)$ but find that this parameterization is more sensitive to
  the value of $\sigma_{\ln M}$.}  We set $\lambda_0 = 35.0$ and $z_0
= 0.25$ as before.

\subsubsection{Line of Sight Cluster Projections}
\label{subsec:projection_effects}

Some \redmap{} objects may in fact be projections of multiple systems
along the line of sight.  We consider only projections of two halos
since projections of three or more halos are significantly less likely
and will therefore have a subdominant impact on our results.  Since
projections typically occur between clusters that are separated by
$\Delta z \lesssim 0.02$, and the width of our redshift bins is
$\Delta z \sim 0.07$, we ignore redshift differences between the two
halos.  In this section we suppress the $z$ dependence of the bias for
notational convenience.

We model the bias of clusters in the presence of projections via
\begin{eqnarray}
\label{eq:projection_effects}
b(\lambda) &=& (1-f) b^{\rm no-proj}(\lambda) + f b^{\rm proj}(\lambda) 
\end{eqnarray}
where $b^{\rm no-proj}(\lambda)$ is given by Eq.~\ref{eq:bias_vs_m_z},
and $b^{\rm proj}(\lambda)$ is the bias of projected clusters of total
observed richness $\lambda$.  The parameter $f$ can be thought of as
the fraction of \redmap{} objects that are projected systems.  To
complete our model, we need to determine an expression for $b^{\rm
  proj}(\lambda)$.

Consider a projected cluster with richness $\lambda$ that is a blend
of two halos of mass $M_1$ and $M_2$.  We assume that the total
richness of the projected system is equal to the sum of the richnesses
of the two projected halos: $\lambda = \lambda_1 + \lambda_2$, where
$\lambda_1$ and $\lambda_2$ are the richnesses of the halos of mass
$M_1$ and $M_2$, respectively.  We define $q=\lambda_1/\lambda$, so
that $q \leq 1.0$.  Without loss of generality, we assume that
$\lambda_1 > \lambda_2$ and so $q\geq 0.5$.

The bias of the projected pair of halos, $b^{\rm proj}(\lambda)$,
should be higher than the bias of either of the individual halos since
projections occur preferentially in high density regions.
Consequently, $b^{\rm proj}(\lambda)\geq b(\lambda_1)$, and thus
\begin{eqnarray}
\label{eq:lower_limit}
b^{\rm proj}(\lambda) >  \int d M_1\ b(M_1)  P(M_1 | q \lambda). 
\end{eqnarray}
Furthermore, $b^{\rm proj}$ should increase as the separation between
$M_1$ and $M_2$ decreases.  In the limit of zero separation, one would
expect the effective bias to be $b(M_1+M_2)$, and hence $b^{\rm
  proj}(\lambda)\leq b(M_1+M_2)$.  Since our model has $M \propto
\lambda^{\alpha}$, then
\begin{eqnarray}
M_2 \sim M_1 \left(\frac{1-q}{q}\right)^{\alpha}.
\end{eqnarray}
We therefore obtain an upper limit to the bias of the projected system
\begin{eqnarray}
\label{eq:upper_limit}
b^{\rm proj}(\lambda) <  \int d M_1\ b\left(M_1\left[ 1 + \left( \frac{1-q}{q} \right)^{\alpha} \right]\right)  P(M_1 | q \lambda).
\end{eqnarray}
We can smoothly interpolate between the lower limit in
Eq.~\ref{eq:lower_limit} and the upper limit in
Eq.~\ref{eq:upper_limit} by introducing a new parameter, $g \in
[0,1]$, that scales the $((1-q)/q)^{\alpha}$ term. 
Our model for the bias including projection effects is then
\begin{multline}
\label{eq:bias_with_proj}
b(\lambda) = (1-f)  \int d M b(M) P( M|\lambda) \\
 + f \int d M_1 b\left(M_1\left[ 1 + g\left( \frac{1-q}{q} \right)^{\alpha} \right]\right)  P(M_1 | q \lambda),
\end{multline}
where $q \in [0.5, 1]$ and $g\in [0, 1]$.  In the limit that $q = 1$,
there is no mass or richness in the less-rich halo and
Eq.~\ref{eq:bias_with_proj} recovers the un-projected expression
(Eq.~\ref{eq:bias_vs_m_z}) as expected.  

We adopt flat priors on $q$ and $g$ across their allowed ranges.  Note
that describing the effects of projections using a single $q$ and a
single $g$ for all clusters is somewhat unrealistic since the values
of these parameters are likely different for different projected
systems.  However, by keeping $q$ and $g$ constant across all
clusters, we are effectively extremizing the effects of projections so
this is a conservative approach.  As for the fraction of projected
clusters $f$, we adopt the prior $f=0.1 \pm 0.04$ utilized in
\citetalias{Simet:2016}.

\subsection{Assembly Bias}
\label{subsec:assembly_bias}

It is commonly assumed that the clustering amplitude of dark matter
halos depends only on the halo mass.  Assembly bias refers to the
dependence of halo clustering on additional properties of the halos,
such as assembly history or concentration.  Assembly bias has been
observed in simulations
\citep[e.g][]{Gao:2005,Wechsler:2006,Jing:2007} and motivated by
theory \citep[e.g.][]{Dalal:2008}.  Recently, \citet{Miyatake:2015}
(hereafter \citetalias{Miyatake:2015}) have found evidence for
assembly bias in a sample of SDSS \redmap{} clusters very similar to
that used in this work.  The results of \citetalias{Miyatake:2015}
suggest that the clustering of \redmap{} clusters depends on $\rmem$,
the mean separation of cluster member galaxies from the cluster
center.

Assembly bias is a potential source of systematic error in the model
we have developed above.  The $b(M)$ from \citet{Tinker:2010} is
calibrated by averaging over all halos of fixed mass $M$ in
simulations.  The \redmap{} catalog, on the other hand, is generated
by selecting on richness, $\lambda$.  Let $\gamma$ represent some
additional property of halos besides mass that affects their
clustering, such as assembly history or concentration.  If the
distribution of $\gamma$ for a richness-selected sample differs from
that of a mass-selected sample, then the clustering amplitude of the
\redmap{} clusters may differ from that predicted by $b(M)$, resulting
in biased parameter constraints.  This effect has been estimated to be
negligible for SDSS clusters \citep{Wu:2008}.  Nevertheless, the
recent detection of assembly bias by \citetalias{Miyatake:2015}
suggests that this effect is significantly stronger in the data than
was originally expected.  Consequently, we explore the the impact of
assembly bias on our mass--richness calibration results in more detail
below.  We also note that in addition to dependence of the halo bias
on assembly history or concentration, the mass--richness relation may
also be dependent on assembly history or halo concentration
\citep{Zentner:2005, Mao:2015}.

\citetalias{Miyatake:2015} consider the dependence of cluster
clustering on the mean distance between the cluster center and the
cluster member galaxies, $\rmem$:
\begin{eqnarray}
\rmem = \sum_i R_i P_i,
\end{eqnarray}
where $R_i$ is the physical separation between the $i$-th member
galaxy and the corresponding halo center, and $P_i$ is the membership
probability of the $i$-th galaxy as determined by the \redmap{}
algorithm.  \citetalias{Miyatake:2015} divide their cluster sample
into two sets with roughly identical richness and redshift
distributions, but with different $\rmem$ distributions.  For both of
these cluster samples, \citetalias{Miyatake:2015} measure the weak
lensing signals around the clusters and fit these measurements with a
halo model.  They find that the one-halo terms for both cluster
samples are consistent within errors (suggesting that the two samples
have similar halo mass distributions), but that the two-halo
clustering amplitudes are different.  \citetalias{Miyatake:2015} also
measure halo clustering directly, and find that the ratio of the
clustering amplitudes for the two samples is consistent with that
found in their weak lensing analysis.  These results suggest that the
clustering of \redmap{} clusters depends not only on the halo masses
of these clusters, but also on $\rmem$ \citep[see also][for further
  discussion]{moreetal06}.

Rather than working with $\rmem$ directly, it is easiest to consider
$\rmem$ normalized by the mean value of $\rmem$ at a given richness
and redshift.  We define
\begin{eqnarray}
\label{eq:delta_def}
\Delta \equiv \frac{\rmem - \langle \rmem | \lambda,z \rangle }{ \langle \rmem | \lambda,z \rangle },
\end{eqnarray}
where $ \langle \rmem | \lambda,z \rangle $ is the mean $\rmem$ for
clusters of richness $\lambda$ and redshift $z$.  We approximate $
\langle \rmem | \lambda,z \rangle $ by fitting a spline to the average
value of $\rmem$ computed across ten bins of richness and five bins of
redshift.

The dependence of halo clustering on $\Delta$ can in principle affect
our analysis in two ways.  First, the mass--richness relationship,
$P(M|\lambda)$, can depend on $\Delta$ (see e.g. \citealt{Mao:2015}).
Since the clustering amplitude depends on the halo mass, any dependence of
the mass--richness relationship on $\Delta$ would induce variation in
the clustering amplitude with $\Delta$.  Secondly, the relationship
between halo mass and clustering, $b(M)$, could directly depend on
$\Delta$.  In principle, both of these effects could be relevant to
the \redmap{} clusters.  However, \citetalias{Miyatake:2015} have
found that their two cluster samples have nearly identical richness
distributions (by construction), very different $\Delta$ distributions
(also by construction), but very nearly identical lensing signals in
the one-halo regime, i.e.  nearly identical mean halo masses.  Their
results suggest (but do not require) that $\Delta$ does not severely
impact the mass--richness relation, but does significantly affect the
clustering amplitude.  That is, we should focus on a model,
$b(M,\Delta)$, in which the halo clustering amplitude is directly
modulated by $\Delta$.

We adopt a linear model for the dependence of bias
on $\Delta$:
\begin{eqnarray}
\label{eq:assembly_bias_model}
b(M, z, \Delta) = b(M,z)(c_0 + c_1 \Delta)
\end{eqnarray}
where $b(M,z)$ is the bias model from \citet{Tinker:2010}.  
Given this dependence of the bias on $\Delta$, Eq.~\ref{eq:bias_vs_m_z}
must be replaced with
\begin{eqnarray}
\label{eq:bias_vs_lambda_delta}
b(\lambda, z, \Delta) &=& \int dM\ P(M|\lambda, z, \Delta) b(M,z,\Delta) \nonumber \\
&=& (c_0 + c_1 \Delta) \int dM\ P(M|\lambda, z) b(M,z).\nonumber \\
\end{eqnarray}
In going from the first to the second line we have assumed that the
mass--richness relationship is independent of $\Delta$.  We note that
\citetalias{Miyatake:2015} found consistent values for the average
mass of clusters in bins of $\Delta$, but this does not require that
there be no dependence of the mass--richness relationship on $\Delta$.
The bias for a richness/redshift bin can then be modeled as in
Eq.~\ref{eq:bias_avg}, except now we include the dependence of the
bias on $\Delta$:
\begin{eqnarray}
\label{eq:bias_avg_delta}
b^{\alpha} = \frac{1}{N_{\alpha}}\sum_i^{N_{\alpha}} b(\lambda_i,z_i,\Delta_i).
\end{eqnarray}

While our model has introduced two new parameters --- $c_0$ and $c_1$
--- in practice these two are related.  Specifically, by construction,
$b(M)$ is the halo bias averaged over all halos of a given mass.
Consequently, one must obtain the \citet{Tinker:2010} mass--bias
relation when averaging over all halos of a given mass:
\begin{eqnarray}
b(M,z) & = & \langle  b \rangle_{M,z} \equiv \int d \Delta \, b(M,z,\Delta) P(\Delta | M, z) , \nonumber \\
\label{eq:satisfy_tinker}
\end{eqnarray}
where we have defined the quantity $\langle b \rangle_{M,z}$, which
represents the bias averaged over the $\Delta$ distribution, $P(\Delta
|M,z)$, at fixed mass and redshift.  Satisfying the above equation
and substituting the ansatz of Eq.~\ref{eq:assembly_bias_model}
yields the constraint
\begin{eqnarray}
\label{eq:constraint}
c_0 + c_1 \langle \Delta \rangle_{M,z} = 1,
\end{eqnarray}
where we have defined
\begin{eqnarray}
\label{eq:meandelta_m_z}
\langle \Delta \rangle_{M,z} \equiv \int d \Delta \, \Delta \, P(\Delta | M, z)
\end{eqnarray}
as the mean $\Delta$ at fixed $M$ and $z$.  

Similarly, we can compute the average bias over the $\Delta$
distribution at fixed richness and redshift:
\begin{eqnarray}
\langle b \rangle_{\lambda, z} & \hspace{-5 pt}  \equiv & \int d \Delta \, b( \lambda, z, \Delta) P(\Delta | \lambda, z) \nonumber \\
 &\hspace{-5 pt}  =& \hspace{-8 pt} \int d \Delta \, (c_0 + c_1 \Delta) P(\Delta | \lambda, z)\int dM P(M|\lambda, z) b(M,z) \nonumber   \\
 &\hspace{-5 pt}  =& (c_0 + c_1 \langle \Delta \rangle_{\lambda,z}) b_0(\lambda,z)  \nonumber \\
 &\hspace{-5 pt} =& \left[ 1 + c_1 \left(\langle \Delta \rangle_{\lambda,z} - \langle \Delta \rangle_{M,z} \right) \right] b_0(\lambda,z).
\label{eq:delta_dist_compare}
\end{eqnarray}
Here, we have defined $b_0(\lambda,z)$ via eq.~\ref{eq:bias_vs_m_z},
i.e. $b_0(\lambda,z)$ is the bias we predict in the absence of
assembly bias.  We have further defined
\begin{eqnarray}
\langle \Delta \rangle_{\lambda,z} \equiv \int d \Delta \, \Delta \,
P(\Delta | \lambda, z)
\end{eqnarray}
in analogy to Eq.~\ref{eq:meandelta_m_z}.  

Notice that from Eq.~\ref{eq:delta_dist_compare} it is clear that if
the $\Delta$ distribution for a richness-selected sample of clusters
is identical to the $\Delta$ distribution of halos at fixed mass, then
$\langle \Delta \rangle_{\lambda,z} = \langle \Delta \rangle_{M,z} $ and the
perturbation term to our prediction exactly cancels out, as it should.
We can, however, simplify our expression just a bit more by noting
that by definition, $\avg{\Delta}_{\lambda,z}=0$.  Consequently,
\begin{eqnarray}
\avg{b}_{\lambda,z} = \left [ 1 - c_1\avg{\Delta}_{M,z} \right] b_0(\lambda,z).
\end{eqnarray}
Thus, we see that the quantity $\langle \Delta \rangle_{M,z}$ governs
the effects of assembly bias on our measurements.

To complete our model for assembly bias, we must determine the values
of $c_1$ and $\langle \Delta \rangle_{M,z}$.  In the absence of a
mass-selected clusters sample, we rely instead on an SZ-selected
cluster sample to estimate $\langle \Delta \rangle_{M,z}$.  We use the
union catalog of SZ-detected clusters from \citep{PlanckXXVII:2015}.
We match the Planck cluster catalog to \redmap{} in order to calculate
$\Delta$ for the Planck clusters.  Note that since all Planck clusters
in the redshift range of our analysis and in the SDSS footprint are
identified by \redmap{}, this results in a complete sample of SZ-selected clusters \citep{rozoetal15}.

\begin{figure}
  \includegraphics[width =\columnwidth]{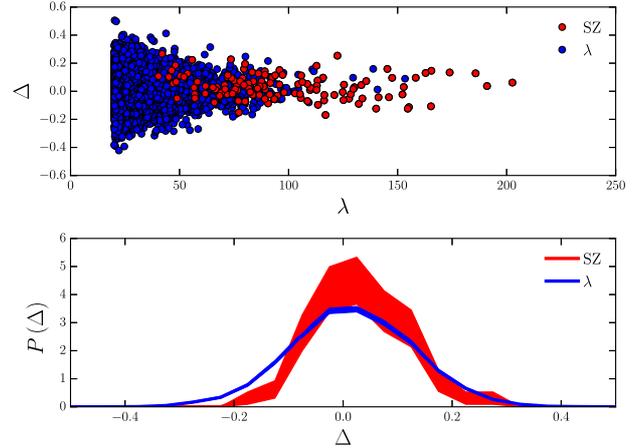}
  \caption{The top panel shows the distribution of clusters in
    richness ($\lambda$) and cluster member concentration ($\Delta$)
    for a richness-selected sample (blue points) and an SZ-selected
    sample from \citet{PlanckXXVII:2015} (red points).  Bottom panel
    shows the distribution of both of these samples as a function of
    $\Delta$; the curve widths indicate the Poisson error bar on the
    measurement of the distribution function.  Comparison of the red
    and blue curves in the bottom panel suggests that
    richness-selected clusters tend to have slightly lower values of
    $\Delta$ than SZ-selected clusters, meaning that richness-selected
    clusters are more compact on average.}
  \label{fig:richness_vs_sz}
\end{figure}

The $\Delta$ distributions of the \redmap{} and Planck-selected
clusters are compared in Fig.~\ref{fig:richness_vs_sz}.  The top panel
of that figure shows the distribution of clusters in richness and
$\Delta$, while the bottom panel shows the distributions as a function
of $\Delta$ alone.  The widths of the curves in the bottom panel are
the Poisson errors for each $\Delta$ bin.  Making the simplifying
assumption that $\langle \Delta \rangle_{M,z}$ is not strongly
dependent on mass or redshift, $\langle \Delta \rangle_{M,z}$ is just
the average value of $\Delta$ for all of the clusters in the
SZ-matched sample.  Performing this average we find $\langle \Delta
\rangle_{M,z} = 0.032\pm 0.007$.  The fact that $\langle \Delta
\rangle_{M,z} > 0$ (measured from the SZ-selected sample) means that
SZ-selected clusters on average have larger values of $\rmem$ than
\redmap{} clusters, i.e. \redmap{} clusters are preferentially more
compact than SZ-selected clusters.

We now turn to estimating $c_1$, the remaining parameter in our model.
Measuring $c_1$ with the SZ-selected sample is difficult because the
small number of SZ-selected clusters means that any correlation
function measurement with these clusters is very noisy.  Instead,
we estimate $c_1$ using the \redmap{} sample in the following way.  In
analogy to Eq.~\ref{eq:assembly_bias_model} we define
\begin{eqnarray}
\label{eq:assembly_bias_model_richness}
b(\lambda, z, \Delta) = \langle b \rangle_{\lambda,z}(d_0 + d_1 \Delta).
\end{eqnarray}
Since $\langle \Delta \rangle_{\lambda,z} = 0$ by definition, we must
have $d_0 = 1$.  By measuring the correlation function in bins of
$\Delta$ we can fit for $d_1$.  We then adopt the approximation
$c_1\approx d_1$.

For the purposes of estimating $d_1$, we divide the cluster sample
into three bins of $\Delta$ and measure the angular clustering of the
clusters in these bins.  We then fit these measurements treating the
bias in each $\Delta$ bin as a free parameter.  The constraints on the
bias parameters as a function of $\Delta$ are shown as the data points
with error bars in Fig.~\ref{fig:b_Delta_fit}.  The position on the
$x$-axis of the data points is the average value of $\Delta$ for that
bin.  We model the bias measurements shown in
Fig.~\ref{fig:b_Delta_fit} with
\begin{eqnarray}
\label{eq:delta_model}
b_{\gamma} = \sum_i^{N_{\gamma}}  b(\lambda_i,z_i) (1 + d_1 \Delta_i),
\end{eqnarray}
where the sum runs over all $N_{\gamma}$ clusters in the $\gamma$-th
$\Delta$ bin and $d_1$ is treated as a free parameter.  For
$b(\lambda_i,z_i)$, we use the best-fit $b(\lambda,z)$ relation from
the bias-only analysis described in \S\ref{subsec:biasonly_param}.

The fitting procedure described above yields a constraint of $d_1 =
3.6\pm0.7$.  We therefore find statistically significant evidence
($5\sigma$) that the clustering of \redmap{} clusters increases with
$\Delta$, in agreement with the results of \citetalias{Miyatake:2015}.
Using this constraint on $d_1$ we can compute a model prediction
(using Eq.~\ref{eq:delta_model}) for the bias of every \redmap{}
cluster.  These predictions, averaged over all clusters in each
$\Delta$ bin are shown as the solid orange bars in
Fig.~\ref{fig:b_Delta_fit}.  The extent of these bars along the
$x$-axis indicates the width of the $\Delta$ bin, while the extent
along the $y$-axis indicates the range allowed by the uncertainty on
$d_1$.  It is clear from the figure that our simple linear model for
the dependence of the bias on $\Delta$ (Eq.~\ref{eq:delta_model})
provides a reasonable fit to the measured biases.  We have performed
similar fits after dividing the \redmap{} clusters into richness bins,
and find no significant evidence that $d_1$ varies with richness.

We now have all the ingredients necessary to compute $c_0$ and $c_1$.
We approximate $c_1 \approx d_1$ and set $c_0 = 1 - c_1 \langle \Delta
\rangle_{M,z}$.  The bias for a richness/redshift bin can then be
computed using Eq.~\ref{eq:bias_avg_delta}.  In computing $c_0$ and
$c_1$ we have made two significant assumptions: first, that the
SZ-selected sample could be used as a reasonable proxy for a
mass-selected sample when computing $\langle \Delta \rangle_{M,z}$,
and second, that we could approximate $c_1 \approx d_1$.  To account
for the systematic error introduced by these assumptions, we take the
approach of allowing $\langle \Delta \rangle_{M,z}$ and $c_1$ to vary
in our analysis across wide, flat priors.  The central values of these
priors are chosen to be $\langle \Delta \rangle_{M,z} = 0.032$ and
$c_1 = 3.6$, i.e. the values measured above.  For the widths of the
priors, we allow $\langle \Delta \rangle_{M,z}$ and $c_1$ to vary
between 50\% and 150\% of their central values.  These priors are
summarized in Table~\ref{tab:mass_richness_parameters}.

Since we find $\langle \Delta \rangle_{M,z} > 0$, \redmap{} clusters
are slightly more concentrated than SZ-selected clusters and
therefore have lower value of $\Delta$ on average.  Since we find $c_1
> 0$, lower values of $\Delta$ correspond to lower clustering biases.
Consequently, for fixed measurements of $w^{\alpha\beta}(\theta)$, a
model that includes the effects of assembly bias is expected to yield
a somewhat higher normalization of the mass--richness relation.
Indeed, as we show in \S\ref{subsec:error_assem}, this is
what we find.

\subsubsection{Evidence for Assembly Bias}

\begin{figure}
  \includegraphics[width=\columnwidth]{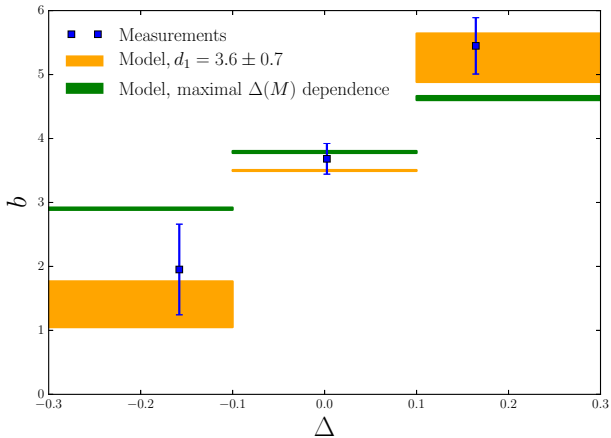}
  \caption{Clustering bias ($b$) as a function of cluster member
    concentration ($\Delta$) for three $\Delta$ bins (blue points).
    The cluster sample used in this plot has been restricted to
    $\lambda \geq 20$.  Orange regions show the best-fit assembly bias
    model prediction (Eq.~\ref{eq:assembly_bias_model}) for the mean
    bias of each $\Delta$ bin.  The vertical extent of the regions
    indicates the range allowed by the uncertainty on the model
    parameter $d_1$.  The green regions represent the biases predicted
    by a model that does not include assembly bias, but instead
    assumes that cluster mass and $\Delta$ are maximally correlated
    (Eq.~\ref{eq:mass_assign}).  The vertical extent of the green
    regions represents the error on the mean bias in this model.  The
    model that includes assembly bias is a better fit to the
    clustering bias measurements than the model that does not include
    assembly bias.}
  \label{fig:b_Delta_fit}
\end{figure}

Given the measured dependence of the bias with $\Delta$ shown in
Fig.~\ref{fig:b_Delta_fit}, an interesting question to ask is whether
or not these measurements provide evidence for assembly bias.  In
principle, the measured dependence of bias with $\Delta$ could be due
to dependence of $\Delta$ on cluster mass.  If higher mass clusters
tended to have higher values of $\Delta$, these clusters would also
have higher biases because $b(M)$ is an increasing function.  If this
were the case, the measured dependence of the bias on $\Delta$ would
not constitute evidence for assembly bias.\footnote{The measurements
  of \citetalias{Miyatake:2015} suggest that $\Delta$ does {\it not}
  depend strongly on cluster mass since the one-halo contributions to
  the weak lensing signals of these two samples are nearly identical.
  Our goal here, however, is to determine whether or not our
  clustering measurements support the assembly bias picture in the
  absence of the mass measurements by \citetalias{Miyatake:2015}.}

The dependence of cluster mass on $\Delta$ must be relatively steep in
order to explain the observed dependence of the bias on $\Delta$.
However, if the dependence of mass on $\Delta$ were too steep, the
resultant scatter in the \redmap{} mass--richness relation would be
larger than the observed scatter.  The most extreme model is that {\it
  all} of the scatter in the mass--richness relation is due to the
dependence of mass on $\Delta$.  If we can show that even this extreme
model cannot explain the observed dependence of bias on $\Delta$, then
this constitutes evidence for assembly bias.  Note that if all of the
scatter in the mass--richness relation {\it were} due to $\Delta$, one
could calibrate a new mass--richness--$\Delta$ relationship with zero
scatter.

To test the extreme model described above, we assign masses to the
\redmap{} clusters using a model that does not include assembly bias,
but instead maximally correlates scatter in the mass--richness
relation with scatter in $\Delta$.  The $i$-th cluster is assigned a
mass using
\begin{eqnarray}
\label{eq:mass_assign}
\ln M_i = \langle \ln M | \lambda_i, z_i \rangle +
\frac{\Delta_i}{\sigma_{\Delta}(\lambda_i)} \sigma_{\ln M}(\lambda_i),
\end{eqnarray}
where we use our best-fit model for $\langle \ln M | \lambda_i, z_i
\rangle$ and $\sigma_{\Delta}(\lambda)$ is the standard deviation of
$\Delta$ as a function of cluster richness.  We compute
$\sigma_{\Delta}(\lambda)$ by binning clusters in richness and
computing $\sigma_{\Delta}$ in each richness bin;
$\sigma_{\Delta}(\lambda)$ is computed at arbitrary richness by linear
interpolation. We assume that the scatter in the mass--richness
relation is described by
\begin{eqnarray}
\sigma_{\ln M}^2(\lambda) 
	& = & \alpha^2\left[ \sigma_{\ln \lambda | M}^2 + (1/\lambda). \right]
\label{eq:lnM_scatter}
\end{eqnarray}
The form of this scatter reflects the population statistics of dark
matter substructures in host halos as found in numerical simulations
\citep[e.g.][]{Boylan-Kolchin:2010, Mao:2015}.  We assume the intrinsic
scatter in mass at fixed richness for high mass clusters (where the $1/\lambda$ term
becomes negligible) is $\sigma_{\ln
  M | \lambda} = \alpha \sigma_{\ln \lambda|M} = 0.3$.  This is a
conservative choice; \citet{redmapperII:2014} and \citet{rozoetal15}
constrain $\sigma_{\ln M | \lambda} \approx 0.25$, including the
Poisson contribution.  By setting $\sigma_{\ln M|\lambda}=0.3$, we are
over-estimating the impact that mass-scatter can have on the
clustering bias prediction.

The mass assignments described by Eq.~\ref{eq:mass_assign} effectively
attributes all of the scatter in the mass--richness relation to
variation in $\Delta$.  With the masses assigned to each cluster using
the no-assembly-bias model of Eq.~\ref{eq:mass_assign}, we compute the
average bias of each $\Delta$ bin in Fig.~\ref{fig:b_Delta_fit} using
the $b(M)$ relation from \citet{Tinker:2010}.  These average biases
and the corresponding error bars are shown as the green solid regions
in Fig.~\ref{fig:b_Delta_fit}.  It is clear from the figure that the
no-assembly-bias model is a worse fit to the bias measurements than
our fiducial assembly bias model
(i.e. Eq.~\ref{eq:assembly_bias_model}).  We find $\chi^2=5.4$ for the
no-assembly-bias model, with a probability to exceed of $6.7\%$ (one
degree of freedom is used to fit for the mean bias).  The measured
bias values can therefore be taken as weak evidence for assembly bias.
This statement is quite conservative since we have assumed a large
value ($\sigma_{\ln M} = 0.3$) for scatter in the $\ln M$--richness
relation in Eq.~\ref{eq:lnM_scatter}, and since we have assumed that
$\Delta$ and cluster mass are maximally correlated.  If the
$\sigma_{\ln M}$ were reduced, or if $\Delta$ and mass are not
maximally correlated, the tension between the no-assembly-bias model
and the measured biases would increase.  Note that if $\Delta$ and
halo mass {\it were} maximally correlated, one could construct a
zero-scatter mass-proxy as some combination of $\lambda$ and $\Delta$;
the implausibility of this scenario suggests that $\Delta$ and halo
mass are almost certainly not maximally correlated.
 
\subsubsection{Calibration Uncertainty in $b(M)$}
\label{subsec:bm_scatter}

\citet{Tinker:2010} find scatter around the $b(M)$ measured across
different simulations of roughly $6\%$.  Such scatter contributes a
fundamental source of uncertainty to the mass--richness constraints we
derive using $b(M)$.  We introduce this uncertainty into our model by
multiplying $b(M)$ by $(1+b_{\rm scaling})$, where $b_{\rm scaling}$
is a new parameter of our model.  We place a Gaussian prior on $b_{\rm
  scaling}$ centered at zero with $\sigma = 0.06$.  In principle,
simply scaling $b(M)$ by $(1+b_{\rm scaling})$ may not fully capture
the cross-simulation variation observed by \citet{Tinker:2010}.
However, this is a reasonable approach given that we do not have
probability distributions for all of the parameters of the
\citet{Tinker:2010} bias model.  Correctly modeling the uncertainty in
$b(M)$ will be an important part of future attempts to use cluster
clustering to constrain mass--observable relations.

\subsection{Likelihood Analysis}

We adopt a Gaussian likelihood for the data given the model:
\begin{eqnarray}
\label{eq:likelihood}
\mathcal{L}(\vec{d} | \vec{p}) &\propto& \exp\left[-\frac{1}{2} \left(\vec{d} - \vec{m}(\vec{p})\right)^T \mathbf{C}^{-1} \left(\vec{d} - \vec{m}(\vec{p})\right) \right], \nonumber \\
\end{eqnarray}
where $\vec{d}$ is the data vector (containing all the cross-richness
correlation function measurements) and $\vec{m}(\vec{p})$ is the model
vector, which is a function of the parameter vector, $\vec{p}$.  We
compute $\mathbf{C}^{-1}$ using the estimator from
\citet{Hartlap:2007}:
\begin{eqnarray}
\label{eq:inv_cov_estimation}
\widehat{\mathbf{C}^{-1}} = \frac{N - d - 2 }{N-1} \mathbf{C}^{-1},
\end{eqnarray}
where $N$ is the number of jackknife regions (in this case $N= 800$),
$d$ is the length of our data vector (in this case $d = 80$) and
$\mathbf{C}$ is the covariance matrix estimated from the jackknifing
procedure.

We remind the reader that we perform two fits to the data: one in
which the bias values themselves are the free parameters, and the
other using a parameterized version of the mass--richness relationship
and the $b(M)$ relation from \citet{Tinker:2010}.  For reference, we
reproduce here the complete model for the bias of the $\alpha$-th
richness/redshift bin:
\begin{multline}
\label{eq:full_bias_model}
b^{\alpha} = \frac{1}{N_{\alpha}} \sum_{i}^{N_{\alpha}} (c_0 + c_1 \Delta_i) b_{\rm scaling} \\
\times \left[ (1-f)  \int d M b(M,z_i) P( M|\lambda_i,z_i)  \right. \\
\left. + f \int d M_1 b\left(M_1\left[ 1 + g\left( \frac{1-q}{q} \right)^{\alpha} \right]\right)  P(M_1 | q \lambda_i,z_i) \right],
\end{multline}
where $P(M|\lambda_i,z_i)$ is given by Eq.~\ref{eq:P_of_lnM}.  As
stated above, we do not model the biases of the low-richness clusters
($5 < \lambda < 20$) with Eq.~\ref{eq:full_bias_model}, but instead
treat the biases of each low richness bin (one for each redshift bin)
as a free parameter.

The analysis of \citet{redmapperII:2014} and \citet{rozoetal15}
suggests that the scatter in the mass--richness relationship is
$\sigma_{\ln M} \sim 0.2-0.3$.  In this analysis, we treat
$\sigma_{\ln M}$ as a free parameter, imposing a flat prior
$\sigma_{\ln M} \in [0.05, 0.5]$.  We find little degeneracy between
$\sigma_{\ln M}$ and $A$ or $\alpha$.  We marginalize over
$\sigma_{\ln M}$ in the constraints presented below.  The priors on
all of our model parameters are summarized in
Table~\ref{tab:mass_richness_parameters}.  These priors, combined with
the likelihood in Eq.~\ref{eq:likelihood} allow us to compute the
posterior on the parameters of our model given the observed data.  We
use a Markov Chain Monte Carlo approach to sample this
multidimensional posterior. We perform the sampling using \verb!emcee!
  \citep{emcee:2013}.  The entire analysis pipeline (from calculating
  $w_M(\theta)$ to sampling the posterior) is implemented in the
  \texttt{CosmoSIS}\footnote{\texttt{https://bitbucket.org/joezuntz/cosmosis/wiki/Home}}
  framework \citep{Zuntz:2015}.

\begin{table*}
  \centering
  \caption{Priors and posteriors on parameters of mass--richness
    parameterization.  The parameters $\ln A$, $\alpha$, $\beta$ and
    $\sigma$ govern the amplitude, slope in richness, slope in mass,
    and scatter of the mass--richness relationship of \redmap{}
    clusters, respectively.  The parameter $f$ is the fraction of
    clusters that are projections, $q$ is the fraction of mass in the
    dominant halo of a projection, and $g$ governs the magnitude of
    projection effects.  The parameter $b_{\rm scaling}$ controls the
    uncertainty in the model $b(M)$ from \citet{Tinker:2010}.  The
    parameters $\langle \Delta \rangle_{M,z}$ and $c_1$ govern the
    assembly bias of \redmap{} clusters.  The parameters $f$, $q$,
    $g$, $b_{\rm scaling}$, $\langle \Delta \rangle_{M,z}$, $c_1$, and
    $\sigma$ are all nuisance parameters that we marginalize over.
    Priors of the form $[A,B]$ are flat with minimum and maximum given
    by $A$ and $B$; priors of the form $A\pm B$ are Gaussian with mean
    given by $A$ and standard deviation given by $B$.  Posteriors are
    represented in terms of the mean and standard deviation.}
  \label{tab:mass_richness_parameters}
  \begin{tabular}{cccc}
    Parameter & Description & Prior & Posterior \\ \hline
    $b^{0,z_0}$ & Bias of $5 < \lambda < 20$, $0.18 < z < 0.26$ clusters &  $[0.2,10.0]$ & $1.85 \pm 0.05$ \\
    $b^{0,z_1}$ & Bias of $5 < \lambda < 20$, $0.26 < z < 0.33$ clusters &  $[0.2,10.0]$ & $1.74 \pm 0.05$ \\
    $\ln (A/M_{\odot})$ & Amplitude of mass--richness relation & $[30.0,35.0]$ & $33.66 \pm 0.18$ \\
    $\alpha$ & Richness scaling of mass--richness relation & $[-0.5, 4.0]$ & $1.18 \pm 0.16$ \\
    $\beta$ & Redshift scaling of mass--richness relation & $[-20.0,20.0]$ & $1.86 \pm 2.4$ \\
    $\sigma$ & Scatter in mass--richness relation & $[0.05, 0.5]$ & --- \\
    $f$ & Projection fraction & $0.10\pm0.04$ & ---\\
    $q$ & Projection effects & $[0.5, 1.0]$ & --- \\
    $g$ & Projection effects & $[0.0, 1.0]$ & --- \\
    $b_{\rm scaling}$ & Uncertainty in $b(M)$ & $0.00\pm0.06$ & --- \\
    $\langle \Delta \rangle_{M,z}$ & Assembly bias & $[0.011, 0.053]$ & ---\\
    $c_1$ & Assembly bias & $[0.95, 3.95]$ & --- \\
  \end{tabular}
\end{table*}

\subsection{Simulations}
\label{subsec:simulations}

In order to validate our approach to measuring the mass--richness
relationship of \redmap{} clusters, we apply the analysis pipeline
that we have developed to a mock data set generated from simulations.
We use an N-body simulation of a flat-$\Lambda$CDM cosmological model
with $\Omega_M = 0.286$, $h_0 = 0.7$, $\Omega_b = 0.047$, $n_S = 0.96$,
and $A_s = 2.1\times10^{-9}$, run with 1400$^3$ particles in a box
1050 Mpc h$^{-1}$ on a side, with the \textsc{L-Gadget} code, a
variant of \textsc{Gadget} \citep{Springel:2005}.  A lightcone was
output from the simulation on the fly, to produce a quarter-sky
simulation over the same redshift range as the data.  We use a halo
catalog generated by the \textsc{Rockstar} halo finder
\citep{Behroozi:2013}, run directly on the dark matter lightcone.

The probability that a halo of mass $M$ has richness $\lambda$,
$P(\lambda | M)$, can be related to our parameterized $P(M|\lambda)$
using Bayes' theorem: $P(\lambda | M) \propto P(M |
\lambda)P(\lambda)$.  The mass--richness relation $P(M | \lambda)$ is
given by Eqs.~\ref{eq:P_of_lnM} and \ref{eq:meanM}.  We determine the
prior $P(\lambda)$ from the real data by fitting a Schechter function
to the observed distribution of $\lambda$.  We randomly draw from
$P(\lambda | M)$ to generate a simulated value of $\lambda$ for each
mock halo.

We choose the parameters $A$, $\alpha$ and $\beta$ of the
mass--richness relation to ensure that the distribution of simulated
clusters in richness and redshift is close to the distribution of real
clusters.  Since the cosmological model used to generate the
simulation used herein is slightly different from the currently
favored $\Lambda$CDM cosmological model, it stands to reason that
matching the observed cluster abundance will require somewhat
different mass--richness parameters than are obtained from our
analysis of the data.  We find that setting $\ln (A/M_{\odot}) =
33.15$, $\alpha = 1.0$, $\beta = 1.0$, and $\sigma_{\ln M} = 0.2$
results in a cluster catalog that has roughly the same number density
of clusters as the SDSS data.

The simulated \redmap{} catalog does not include the effects of
line-of-sight projections of clusters, because we use a pure halo
catalog generated using the three-dimensional positions of the dark
matter particles.  As a result, two halos that happen to be close
together on the sky will never be lumped into the same simulated
\redmap{} object.  For this reason, when analyzing the simulation data
we fix $f=0$.  Furthermore, while the simulations include the effects
of assembly bias in the halos, without a more comprehensive modeling
approach, we do not know an appropriate value of $\Delta$ to assign to
each cluster.  We therefore fix $\langle \Delta \rangle_{M,z} = 0$ and
$c_1=0$ when analyzing the simulated \redmap{} catalog.

\begin{figure}
\includegraphics[width = \columnwidth]{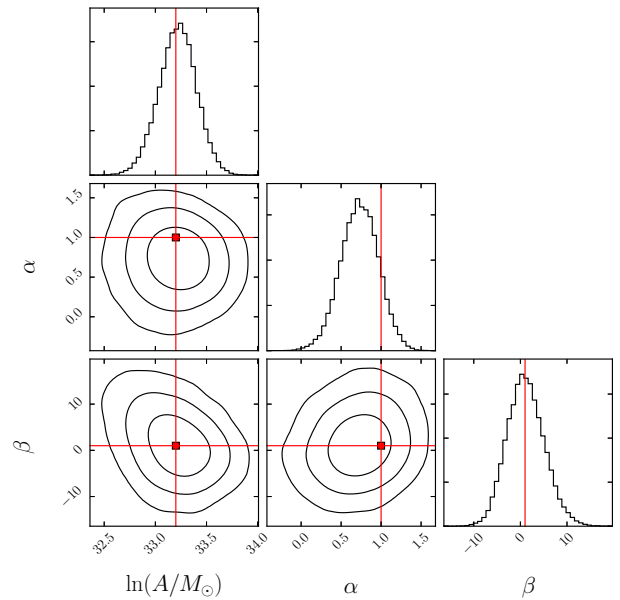}
\caption{Posteriors on the parameters of the mass--richness relation
  resulting from the analysis of a simulated cluster catalog generated
  from N-body simulations.  Contour lines enclose 68\%, 95\% and
  99.7\% of the probability density.  Input values are shown by the
  red lines.  Our analysis successfully recovers the input parameters of the
  mass--richness relation.}
\label{fig:sim_results}
\end{figure}

We apply the same analysis pipeline to the simulated data as to the
real data.  This includes the process of pair counting, jackknifing to
determine the covariance matrix, and fitting for the mass--richness
parameters.  When computing $w_M(\theta)$ for the analysis of the
simulated data we use the same cosmological model that was used to
generate the N-body simulations.  

Fig.~\ref{fig:sim_results} shows the constraints obtained on the
mass--richness parameters by applying our pipeline to the simulated
data.  The input mass--richness parameters are shown as the red lines,
and contour lines represent $1\sigma$, $2\sigma$, etc.  As evidenced
by the figure, our analysis appears to correctly extract the input
model parameters.  We find $\chi^2/\mbox{d.o.f.}=70.7/73$, suggesting
that the model is an excellent fit to the data.  Furthermore, we find
that the error bars on the extracted model parameters are similar to
those found in our analysis of the real data.

\section{Results}
\label{sec:results}

Figures showing the measured angular correlation functions and the
best fit models are collected in Appendix~\ref{app:data},
Figs.~\ref{fig:data_z1} and \ref{fig:data_z2}.  The best-fitting
models for the bias-only parameterization are shown in red, while the
best-fitting mass--richness models are shown in green.  The results of
these two fits are very similar (in many cases the two curves are
indistinguishable by eye). This suggests that the mass--richness
parameterization is well matched to the true clustering biases of the
\redmap{} clusters.

We find somewhat low values of $\chi^2$, with
$\chi^2/\mbox{d.o.f.}=57.1/72$ for the bias-only model, and
$\chi^2/\mbox{d.o.f.}=57.5/75$ for the mass--richness model.\footnote{
  When calculating the d.o.f. for the mass--richness model, we do not
  include the various systematic parameters that have small impacts on
  our model predictions.}  The probabilities for $\chi^2$ to be less
than the quoted values are roughly $10\%$ and $6\%$, respectively. We
caution, however, that since the covariance matrix is estimated using
a Jackknife approach, there are noise fluctuations in the covariance
matrix that can increase or reduce the recovered scatter.  Roughly
speaking, if our error bars on the correlation function measurements
were reduced by $\sim 10\%$, we would have $\chi^2/\mbox{d.o.f.} \sim
1$.  As we show below, our error bars on the parameters of the
mass--richness relation are dominated by uncertainty in the model
parameters.  A $10\%$ reduction in the measurement errors would
therefore have little impact on our results.

\begin{table}
\centering
\caption{Marginalized constraints on the biases of \redmap{} clusters
  in bins of richness and redshift.}
\label{tab:bias_constraints}
\begin{tabular}{ccc}
Richness range &  $0.1 < z < 0.26$ & $0.26 < z < 0.33$ \\
\hline
$5 < \lambda < 20$ & $b = 1.85 \pm 0.06$  & $b = 1.74 \pm 0.05$ \\
$20 < \lambda < 28$ & $b = 2.8 \pm 0.2$ & $b = 3.0 \pm 0.2$ \\
$28 < \lambda < 41$ & $b = 3.1 \pm 0.3$  & $b = 3.7 \pm 0.3$ \\
$41 < \lambda < \infty$  & $b = 4.6 \pm 0.3$  & $b = 5.1 \pm 0.4$ 
\end{tabular}
\end{table}

Table \ref{tab:bias_constraints} summarizes the constraints on the
clustering biases for each individual richness/redshift bin.  As
expected, the bias increases steadily with increasing richness.

\begin{figure}
\includegraphics[width = \columnwidth]{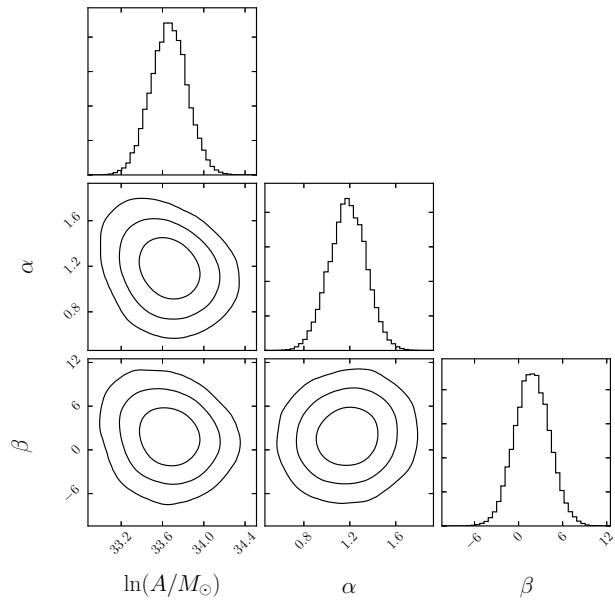}
\caption{Constraints on the parameters of the mass--richness
  relationship obtained from a fit to the measured angular correlation
  functions.  Contour lines enclose 68\%, 95\% and 99.7\%
  of the probability density. }
\label{fig:mass_parameter_constraints}
\end{figure}

Fig.~\ref{fig:mass_parameter_constraints} shows the constraints on the
parameters of the mass--richness relation derived from the angular
clustering measurements.\footnote{Contour plots were made using
  \texttt{corner.py} \citep{corner:2016}.}  The value of $\ln A$ ---
the amplitude parameter of the mass--richness relationship --- is
constrained to be $\ln (A/M_{\odot}) = 33.66 \pm 0.18$, corresponding
to an 18\% mass constraint.  The slope of the mass--richness
relationship is constrained to be $\alpha = 1.18 \pm 0.16$.  The
scaling with redshift is constrained to be $\beta =1.86 \pm 2.4$; the
weakness of this constraint is due in part to our use of only two
redshift bins and the fact that our cluster sample is narrowly
distributed in redshift.  These constraints are summarized in the
`Posterior' column of Table~\ref{tab:mass_richness_parameters}.

Our calibration can be compared to the weak lensing mass estimator of
\citetalias{Simet:2016}.  We emphasize that despite common
  authors between this work and \citetalias{Simet:2016}, no comparison
  between the two analyses was performed before the analyses pipelines
  were finalized.  Our decision to focus on the NGC exclusively did
  come after a comparison with \citetalias{Simet:2016} had occurred,
  though we believe we are well justified in our choice.

\citetalias{Simet:2016} provides the current best weak lensing mass
calibration of \redmap{} clusters, and they discuss in detail how
their mass calibration compares to others in the literature.  While
the constraints on the slope of the mass--richness relation are in
good agreement between the two works, there is mild tension on the
recovered amplitude assuming our fiducial cosmology, with the
clustering-derived constraints preferring a somewhat higher amplitude.
At the richness pivot point used in this work, $\lambda_0 = 35$,
\citetalias{Simet:2016} predicts $\log_{10} (M/M_{\odot})
=14.43\pm0.04$, while our constraints yield $\log_{10} (M/M_{\odot}) =
14.62\pm 0.08$; this discrepancy corresponds to roughly $2.2\sigma$
tension.  Because the clustering analysis prefers a slightly lower
value of $\alpha$ than the weak lensing analysis, this tension is
reduced for $\lambda > 35$.  For $\lambda < 35$, the tension remains
roughly the same since the error bars increase as one moves away from
the $\lambda$ pivot point.  We postpone a discussion of the origin of
this tension, and a more detailed comparison of the two results, to
future work.  That is, our current results reflect our best
understanding of both the weak lensing and clustering analysis prior
to the comparison of the two works.

We have also considered a power-law bias-richness relation as per
Eq.~\ref{eq:bias_vs_lam_z}.  This model has five parameters:
$A_{\lambda}$, $\alpha_{\lambda}$, $\beta_{\lambda}$ and two bias parameters
for $5 < \lambda < 20$ clusters in the two redshift bins.  Fitting
this model to the data, we find $A_{\lambda} = 3.54\pm0.13$,
$\alpha_{\lambda} = 0.57\pm0.08$ and $\beta_{\lambda} = 2.0\pm1.1$ with
$\chi^2/{\rm d.o.f.} = 57.7/75$.

\section{Characterizing the Uncertainty in the Mass--Richness Relation}
\label{sec:systematics}

The uncertainties on the mass--richness parameters presented in
Fig.~\ref{fig:mass_parameter_constraints} and
Table~\ref{tab:mass_richness_parameters} receive contributions from
several sources: statistical noise in the measurement of the angular
correlation functions, uncertainty on the parameters of the
line-of-sight projection model, uncertainty on the parameters of the
assembly bias model, and scatter around the $b(M)$ relation from
\citet{Tinker:2010}.  In addition to these sources of scatter that are
explicitly included in our model, it is also possible that our
measurement is affected by systematic biases that we have not modeled.
Below, we determine the approximate contributions of both modeled
sources of scatter and unmodeled systematic errors to the uncertainty
budget of our parameter constraints.

\subsection{Statistical Noise in the Correlation Function Measurement}
\label{subsec:error_cov}

We first investigate how statistical noise in the measurement of
$w^{\alpha\beta}(\theta)$ contributes to our parameter uncertainties.
The contribution of statistical noise to our parameter uncertainties
can be determined by setting the parameters of the line-of-sight
projection model, the parameters of the assembly bias model, and
$b_{\rm scaling}$ to constant values rather than allowing these
parameters to vary.  Any uncertainty in our parameter constraints that
remains after fixing these model parameters must be due to measurement
noise alone.

For the purposes of this test, we fix $f=0.1$, $q = 0.75$, $g = 0.5$,
$\langle \Delta \rangle_{M,z} = 0.032$, $c_1 = 3.6$ and $b_{\rm
  scaling} = 0$, corresponding to the central values of the priors on
these parameters.  The results of this section are relatively
insensitive to the precise values that we choose here.  Upon
re-fitting this constrained model to the data, we find that the
posterior on $\ln A$ is significantly tighter than that obtained in
our fiducial analysis.  In other words, the uncertainty on the
amplitude of the mass--richness relation is dominated by uncertainty
on the model parameters that we marginalize over.  In the absence of
uncertainty on these model parameters, the error on $\ln A$ would be
only $\sim 0.07$ instead of $\sim 0.18$.  Statistical noise in the
measurement of $w^{\alpha\beta}(\theta)$ contributes only $\sim 10\%$
of the variance of our baseline constraint on $\ln A$.

On the other hand, we find that statistical measurement error
contributes $\sim 70\%$ of the variance of our constraints on $\alpha$
and $\beta$.  The uncertainties on these parameters would be
significantly reduced with a larger cluster catalog distributed over a
wider redshift range.

\subsection{Scatter around $b(M)$ from \citet{Tinker:2010}}
\label{subsec:b_of_m_scatter}

Our fiducial analysis assumes a scatter of 6\% around the the $b(M)$
model from \citet{Tinker:2010}, parameterized with $b_{\rm scaling}$.
To measure the contribution of this source of scatter to our parameter
uncertainties, we consider a model that allows $b_{\rm scaling}$ to
vary with the fiducial prior, but that fixes all other model
parameters to the values given in \S\ref{subsec:error_cov}.  Comparing
the posteriors obtained from fitting this model to the data to those
obtained in \S\ref{subsec:error_cov} provides a measure of the
contribution of uncertainty on $b_{\rm scaling}$ to our final
parameter uncertainties.

We find that uncertainty on $b_{\rm scaling}$ dominates the
uncertainty on $\ln A$, contributing roughly $60\%$ of the variance of
our fiducial constraint on this parameter.  In other words, our
constraint on the normalization of the mass--richness relation for
\redmap{} clusters is currently limited by uncertainty in the $b(M)$
relation.  In contrast, the posteriors on $\alpha$ and $\beta$ are
less affected by uncertainty on $b_{\rm scaling}$; uncertainty on
$b_{\rm scaling}$ contributes only $\sim 30\%$ of the variance of our
constraints on these parameters.

\subsection{Uncertainty on Assembly Bias Parameters}
\label{subsec:error_assem}

To evaluate the contribution of uncertainty on the assembly bias
parameters $c_1$ and $\langle \Delta \rangle_{M,z}$ to our constraints
on the mass--richness parameters, we fit a model to the data that
allows the assembly bias parameters to vary over their fiducial
priors, but that keeps the other model parameters fixed to the values
in \S\ref{subsec:error_cov}.  We find that uncertainty on the assembly
bias parameters contributes roughly $30\%$ of the variance of our
baseline constraint on $\ln A$, but contributes negligibly to the
variance of the constraints on $\alpha$ and $\beta$.  Uncertainty on
the assembly bias parameters therefore contributes a level of
uncertainty on the normalization of the mass--richness relation that
is greater than that of measurement uncertainty.

We also evaluate how much the inclusion of assembly bias effects
shifts our parameter constraints relative to a model that does not
include these effects.  We fit the data with a model that has $\langle
\Delta \rangle_{M,z} = 0.0$ and $c_1 = 0$, but otherwise is identical
to the baseline analysis.  The results of this fit suggest that the
inclusion of assembly bias in our analysis causes our best-fit $\ln A$
to increase by roughly 15\%, corresponding to $0.8\sigma$.  The
direction of this shift is as anticipated in
\S\ref{subsec:assembly_bias}.  The inclusion of assembly bias effects
has a smaller impact on the best-fit values of $\alpha$ and $\beta$.

\subsection{Uncertainty on Line-of-sight Projection Parameters}
\label{subsec:error_proj}

We follow a similar approach to that outlined in
\S\ref{subsec:error_assem} to estimate the contribution of uncertainty
on the line-of-sight projection parameters $f$, $q$ and $g$ to our
mass--richness parameter uncertainties.  We fit a model to the data
that allows the projection parameters to vary over the priors in
Table~\ref{tab:mass_richness_parameters}, but which has the assembly
bias and $b_{\rm scaling}$ parameters fixed to the same values as in
\S\ref{subsec:error_cov}.  Comparing the parameter uncertainties
obtained using this constrained model to the parameter uncertainties
obtained in \S\ref{subsec:error_cov} suggests that uncertainty on the
projection parameters contributes less than $10\%$ of the variance of
$\ln A$, $\alpha$, and $\beta$.

We can also ask how much line-of-sight projections shift our parameter
constraints relative to an analysis that does not account for
projections.  To do this, we fit a model to the data that has $f=0.0$,
but otherwise is identical to the baseline model.  Analyzing the data
with this model reveals that including line-of-sight projections in
our analysis shifts the maximum likelihood values of $\ln A$,
$\alpha$, and $\beta$ by less than $0.1\sigma$.

\subsection{Systematics in $w^{\alpha\beta}(\theta)$ Measurement}
\label{subsec:err_sysweights}

As described in \S\ref{subsec:sys_correction}, we apply weights to the
\redmap{} random catalog to correct for an observed dependence of the
cluster density on the $g$-band sky flux.  A similar variation in the
cluster density is also observed with the $E(B-V)$ dust extinction
maps from \citet{Schlegel:1998} (see
Fig.~\ref{fig:density_vs_systematic}).  To determine the level of
systematic error introduced into our measurements by only correcting
for the correlation between cluster density and $g$-band sky flux, we
repeat the correlation function measurement and fitting using a random
catalog weighted instead by $E(B-V)$ dust extinction.

We find that the application of the $E(B-V)$ weights causes our
best-fit $\ln A$, $\alpha$, and $\beta$ to change by less than
$0.05\sigma$.  This suggests that the systematic error introduced by
only correcting for variations in cluster density with $i$-band PSF is
small compared to our error bars.

As mentioned above, the results presented thus far were derived using
only the NGC region of SDSS.  We have also applied our clustering
analysis to the SGC region.  We find some tension between the
clustering measurements in both regions, with the SGC preferring
somewhat higher values of $\ln A$ and somewhat lower values of
$\alpha$ than the NGC.  Considering the bias-only parameterization, we
find the amplitude of the bias--mass relation to be consistent between
the two regions.  However, the constraints on $\alpha_{\lambda}$ for
the two regions differ by $2.1\sigma$.  There is also roughly
$2.4\sigma$ tension in the biases of the clusters in the
richness/redshift bin $5 < \lambda < 20$, $0.18 < z < 0.26$ between
the two regions.  There is no significant tension in the
$\beta_{\lambda}$ constraints between the two regions.  When
considering constraints on the mass--richness parameterization, the
tension between NGC and SGC is reduced because the marginalization
over $b_{\rm scaling}$ and the assembly bias parameters effectively
increases the error bars on the parameters.  Tension in the $\ln A$
constraints from the two regions is only $0.3\sigma$, while tension on
$\alpha$ is $2\sigma$.

\begin{figure}
\includegraphics[width = \columnwidth]{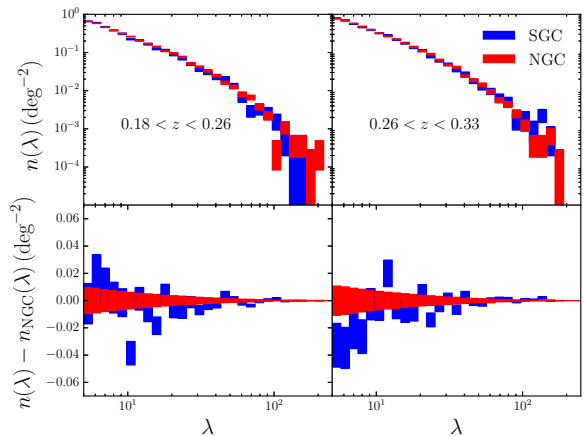}
\caption{Cluster density as a function of richness for the NGC and SGC
  regions.  We find some evidence for tension in the abundance of
  clusters between the NGC and SGC; this tension is driven by
  low-redshift clusters with $\lambda \sim 10$.}
\label{fig:density_vs_richness}
\end{figure}

If the clustering of \redmap{} clusters is different in the NGC and
the SGC, one might expect the abundance of \redmap{} clusters to also
be different in the NGC and the SGC.
Fig.~\ref{fig:density_vs_richness} shows a comparison of the cluster
density as a function of richness between the NGC and the SGC.  The
error bars in the figure only include the Poisson contribution and not
sample variance.  In general, the agreement between the two regions
with respect to cluster abundance appears to be quite good.  However,
we find $\chi^2/\mbox{d.o.f.}=43.7/26$ for the low redshift bin.  The
high value of $\chi^2$ is driven almost entirely by clusters with
$\lambda \sim 10$.  For the high redshift bin, we find
$\chi^2/\mbox{d.o.f.}=34.8/26$.  Note, though, that by only including
the Poisson contribution to the error bars, we have over estimated the
tension between the NGC and SGC in this comparison.

In summary, we find some evidence for tension between the
clustering-derived parameter constraints in the NGC and SGC.  This
tension is most pronounced for our constraints on $\alpha$: the NGC
prefers a higher value of $\alpha$ than the SGC by roughly $2\sigma$.
Tension between the two regions with respect to the other
mass--richness parameters is consistent with noise.  There is also
some evidence for tension in the abundance of low redshift, $\lambda
\sim 10$ clusters between the two regions. We note that
\citet{Tojeiro:2014} also found evidence for tension between the NGC
and SGC at low redshift ($z< 0.25$) in their measurements of the
clustering of galaxies identified in SDSS.  Since the origin of this
tension is not well understood at this time, we refrain from combining
the parameter constraints obtained from the NGC and SGC, and instead
present only the results for the NGC.  This choice also has the
advantage that it makes comparison to the weak lensing mass
constraints from \citetalias{Simet:2016} more straightforward since
the \citetalias{Simet:2016} analysis was only applied to clusters in
the NGC.

\subsection{Jackknife Covariance Estimation}
\label{subsec:err_jack}

Our jackknife estimate of the covariance matrix does not show any
obvious evidence for systematic errors.  However, we feel it necessary
to confirm that our covariance estimate is robust with respect to
the choice of jackknife regions.  To do this, we repeat the
jackknifing procedure with 10\% fewer jackknife regions (720 instead
of $\sim 800$).  Re-fitting the correlation function measurements for
the mass--richness parameters using the modified covariance estimate
leads to small shift in the best-fit mass--richness parameters, all
well within the error bars.  The largest shift observed is in the
best-fit value of $\alpha$, which changes by $0.23\sigma$.  Given the
smallness of these shifts compared to the error bars, any uncertainty
in our mass--richness constraints introduced by the choice of
jackknife regions can be safely ignored.

\section{Cosmology Dependence}
\label{sec:cosmology_dependence}

The results presented above have assumed a fixed cosmological model.
In this section, we vary the cosmological parameters and re-fit the
data to determine how the mass--richness relation varies with
cosmology.  We focus here on $\Omega_M$, $A_S$ and $h_0$.  Other
cosmological parameters are expected to have a subdominant effect on
the predicted correlation function.  

We compute the maximum likelihood mass--richness parameters at 12
points in $(\Omega_M, A_S, h)$ parameter space.  These points are
chosen to be within 10\% of the fiducial values of $\Omega_M$, $A_S$
and $h_0$ since \citet{PlanckXIII:2015} constrains these parameters to
better than 10\%.  Furthermore, restricting $\Omega_M$, $A_S$ and
$h_0$ to a narrow interval ensures that the dependence of the
mass--richness relation on these parameters is close to linear over
the sampled range.  For each point $(\Omega_M, A_S, h)$ we determine
the maximum likelihood values of $\ln A$, $\alpha$, and $\beta$ given
our correlation function measurements.  Using linear regression, we
then compute the dependence of $\ln A$, $\alpha$, and $\beta$ on
$\Omega_M$, $A_S$ and $h_0$.

For the amplitude parameter, $\ln A$, we find a best fit linear model:
\begin{eqnarray}
\ln (A/M_{\odot}) &=& 33.68 + 2.9\delta \Omega_M + 1.0\delta A_s + 1.8\delta h
\end{eqnarray}
where we have defined
\begin{eqnarray}
\delta \Omega_M &=& \frac{\Omega_M - 0.3089}{0.3089} \\
\delta A_S &=& \frac{A_S - 2.141\times10^{-9}}{2.141\times10^{-9}} \\
\delta h &=& \frac{h - 0.6774}{0.6774}
\end{eqnarray}
as the fractional departures of $\Omega_M$ and $A_S$ from the values
assumed in our baseline analysis.  We find that 10\% changes in
$\Omega_M$, $A_S$ and $h_0$ result in $1.6\sigma$, $0.6\sigma$ and
  $1.0\sigma$ shifts in $\ln A$, respectively.  Note that $\Omega_M$,
  $A_S$ and $h_0$ are constrained to $\sim 2\%$ by
  \citet{PlanckXIII:2015}; the variation in $\ln A$ as a function of
$\Omega_M$, $A_s$ and $h_0$ over the range allowed by Planck is
therefore below the uncertainty on $\ln A$.

We find that $\alpha$ is most senstive to $\Omega_M$ and $h_0$, with no
significantly measured dependence on $A_S$ over the range of $A_S$
values considered.  Our best fit relation is 
\begin{eqnarray}
\alpha &=& 1.17 - 0.6 \delta \Omega_M - 0.5 \delta h.
\end{eqnarray}
Changing $\Omega_M$ and $h_0$ by 10\% therefore changes $\alpha$ by
$0.4\sigma$ and $0.3\sigma$, respectively.  Any dependence of $\alpha$
on $A_S$ is significantly below our error bar on $\alpha$.

We find the $\beta$ is most sensitive to $h_0$, with a best-fit linear
relationship described by
\begin{eqnarray}
\beta &=& 1.8 + 1.1\delta_h.
\end{eqnarray}
A 10\% change in $h_0$ therefore results in a $0.05\sigma$ change to
$\beta$.  The dependence of $\beta$ on $\Omega_M$ and $A_S$ is well
below our error bar on $\beta$ and we do not report it here.

\section{Discussion}
\label{sec:discussion}

We have measured the angular correlation function of \redmap{}
clusters identified in SDSS data.  By fitting models to these
measurements, we have extracted constraints on several parameters
describing the clustering biases and mass--richness relationship of
\redmap{} clusters.  Our constraints on the biases of \redmap{}
clusters in bins of richness and redshift are shown in
Table~\ref{tab:bias_constraints}.  Our constraints on the
mass--richness relationship for \redmap{} clusters are shown in
Fig.~\ref{fig:mass_parameter_constraints} and summarized in
Table~\ref{tab:mass_richness_parameters}.

We measured the correlation between clustering bias and $\Delta$,
where $\Delta$ is related to the concentration of member galaxies and
is defined in Eq.~\ref{eq:delta_def}.  These measurements support the
results of \citetalias{Miyatake:2015} and \citet{More:2016}.
Expanding on those works, we have also measured the slope of the
bias-$\Delta$ relationship.  Furthermore, our clustering measurements
provide evidence for assembly bias independent of the weak lensing
measurements of \citetalias{Miyatake:2015}.  Even if one assumes that
$\Delta$ and mass are maximally correlated, the clustering
measurements shown in Fig.~\ref{fig:b_Delta_fit} cannot be explained
without invoking assembly bias.  Finally, we find that the inclusion
of assembly bias effects in our analysis increases the best-fit
amplitude of the mass--richness relationship by $\sim 15\%$,
reflecting the fact that the signal seen by \citetalias{Miyatake:2015}
is significantly larger than was expected {\it a priori}.

In the absence of systematic sources of uncertainty, our analysis
constrains the amplitude of the mass--richness relation to $\sim 7\%$.
Including systematic sources of uncertainty, however, degrades our
constraint on the amplitude of the mass--richness relation to 18\%.
The dominant source of systematic uncertainty is scatter around the
$b(M)$ relation from \citet{Tinker:2010}, which makes up roughly 60\%
of the variance in our final constraint on $\ln A$.  Uncertainty on
the assembly bias parameters is the next largest source of systematic
error affecting our constraint on $\ln A$, and contributes a level of
uncertainty on this parameter that is greater than the uncertainty due
to measurement noise.  We find that line-of-sight projection effects
have a negligible impact on our parameter constraints.  On the other
hand, our constraints on the slope parameters of the mass--richness
relation --- $\alpha$ and $\beta$ --- are not as strongly affected by
systematic uncertainties and are currently statistics limited.

We emphasize that future measurements of cluster clustering with e.g. the
Dark Energy Survey can significantly reduce the statistical
uncertainty on the measured correlation functions, and enable the
assembly bias parameters to be measured more tightly.  In particular,
we note that the larger the volume probed, the lower the errors, so
cluster-clustering is particularly well suited for calibrating the
mass--richness relation of high redshift clusters, exactly where the
statistical and systematic uncertainty in weak lensing mass
calibration is greatest.  Critically, the current errors in the
recovered mass--richness relation are systematic dominated, with the
dominant systematic being theoretical uncertainty in the calibration
of the bias--mass relation of dark matter halos.  Thus, unlike weak
lensing mass calibration efforts, a path towards improved mass
calibration from cluster clustering is very straight forward: one just
needs to implement a simulation program geared toward improving the
calibration of halo bias.  The prospect of improved data sets, reduced
statistical errors, and a straight forward path towards reducing
systematic uncertainties all bode well for the future of
cluster-clustering as a method for calibrating the mass--richness
relation of galaxy clusters.

\section*{Acknowldegements}

This work received partial support from the U.S.\ Department of Energy
under contract number DE-AC02-76SF00515 and from the National Science
Foundation under NSF-AST-1211838.  EB and BJ are partially supported
by the US Department of Energy grant DE-SC0007901.  We thank Matthew
Becker and Michael Busha for their contributions to the N-body
simulations used in this work, and Yao-Yuan Mao for helpful
discussions about assembly bias.  We thank Mike Jarvis for developing
\texttt{treecorr} and for help with running it.  We thank Melanie
Simet, Rachel Mandelbaum, and Gary Bernstein for useful discussions
related to this work.

\bibliographystyle{mnras}
\bibliography{ms}

\appendix

\section{Measurements and Best Fit Models}
\label{app:data}

\begin{figure*}
\includegraphics[scale = 0.9]{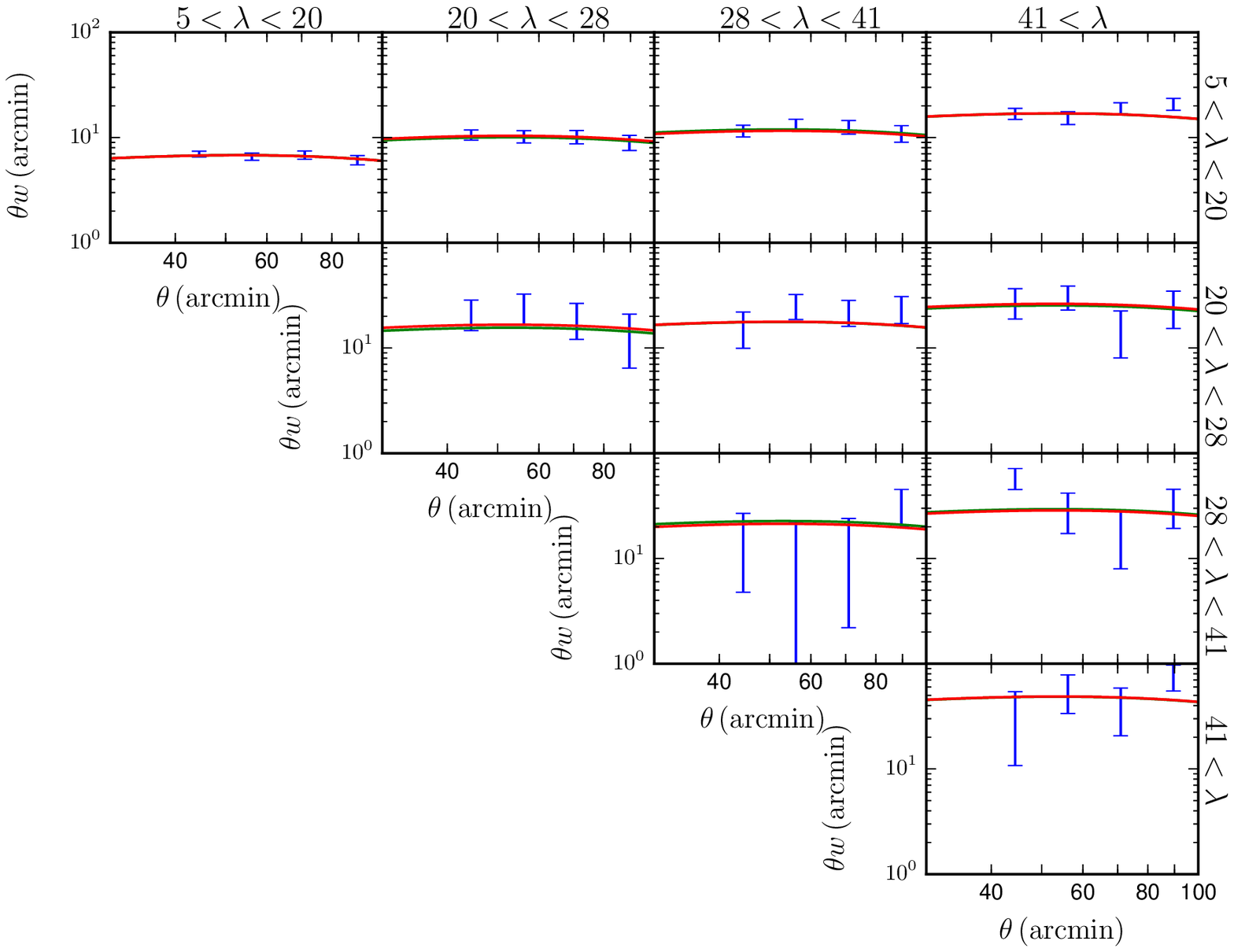}
\caption{The angular cross-correlation function, $\theta
  w^{\alpha\beta}(\theta)$ between \redmap{} clusters in several
  different richnesses bins for the redshift interval $0.18 < z <
  0.26$.  The richness bins are indicated on the top and right sides
  of the figure.  Error bars are the diagonal elements of the
  covariance matrix.  The red curves shows the best-fitting model when
  each richness bin is assigned a free bias parameter, while the green
  curves are obtained through modeling of the mass--richness relation
  (in many cases, the green and red curves are indistinguishable).}
\label{fig:data_z1}
\end{figure*}

\begin{figure*}
\includegraphics[scale = 0.9]{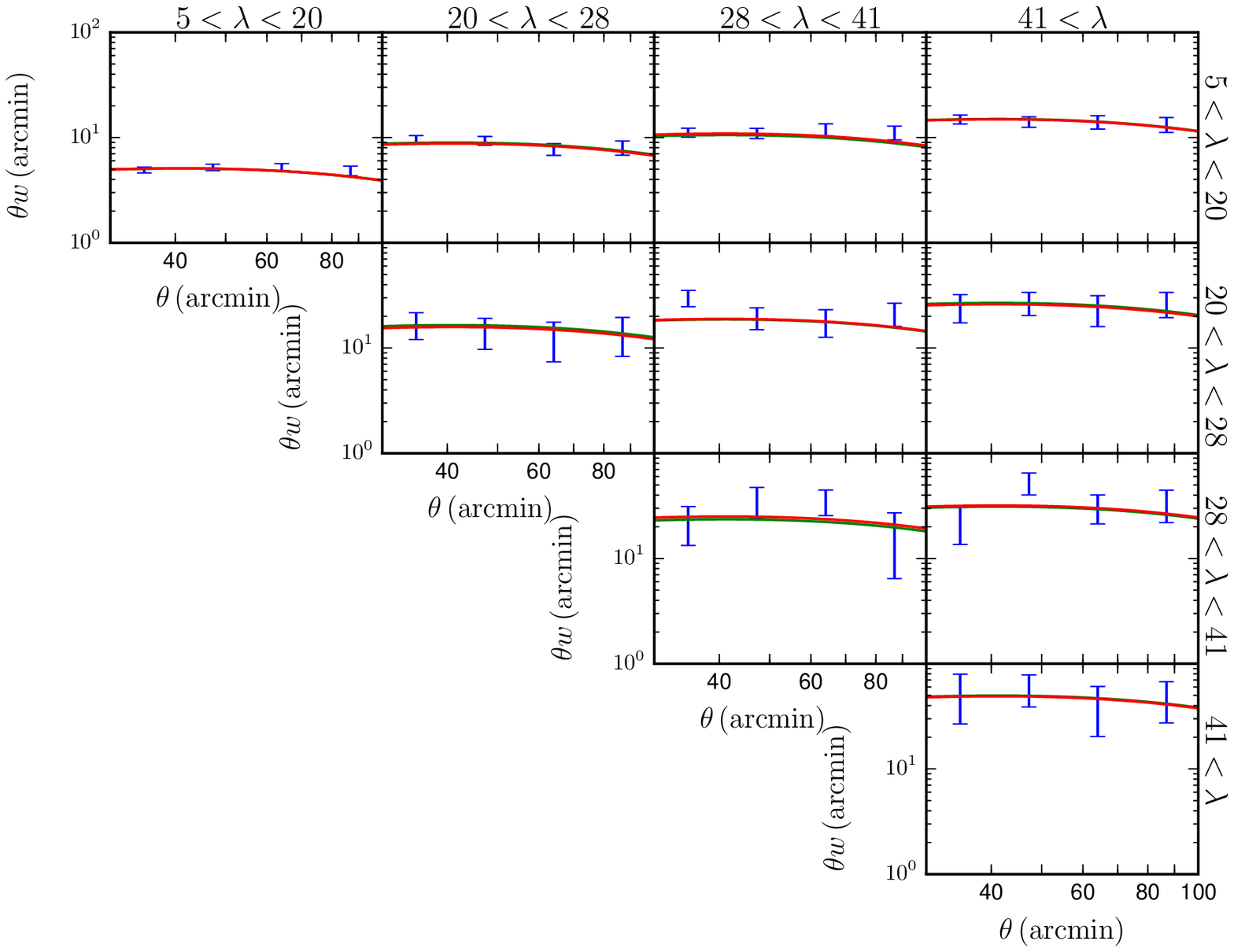}
\caption{Same as Fig. \ref{fig:data_z1}, but for \redmap{} clusters
  with $0.26 < z < 0.33$.}
\label{fig:data_z2}
\end{figure*}

Figures~\ref{fig:data_z1} and \ref{fig:data_z2} shows the angular
correlation function for \redmap\ clusters for two different redshift
bins, and our best fit model both when adopting a free bias parameter
for each richness bin (red) and when modeling the mass--richness
relation (green).

% Don't change these lines
\bsp	% typesetting comment
\label{lastpage}

\end{document}